\documentclass[aps,superscriptaddress,12pt,nofootinbib,a4paper,floatfix]{revtex4} 
\usepackage{graphicx}

\begin{document}

\title{Model-independent constraints on $\mathbf{\Delta F=2}$
  operators\\ and the
  scale of New Physics
  \vspace*{0.5cm}
  \begin{figure}[htb!]
 \begin{center}
 \includegraphics[width=0.13\textwidth]{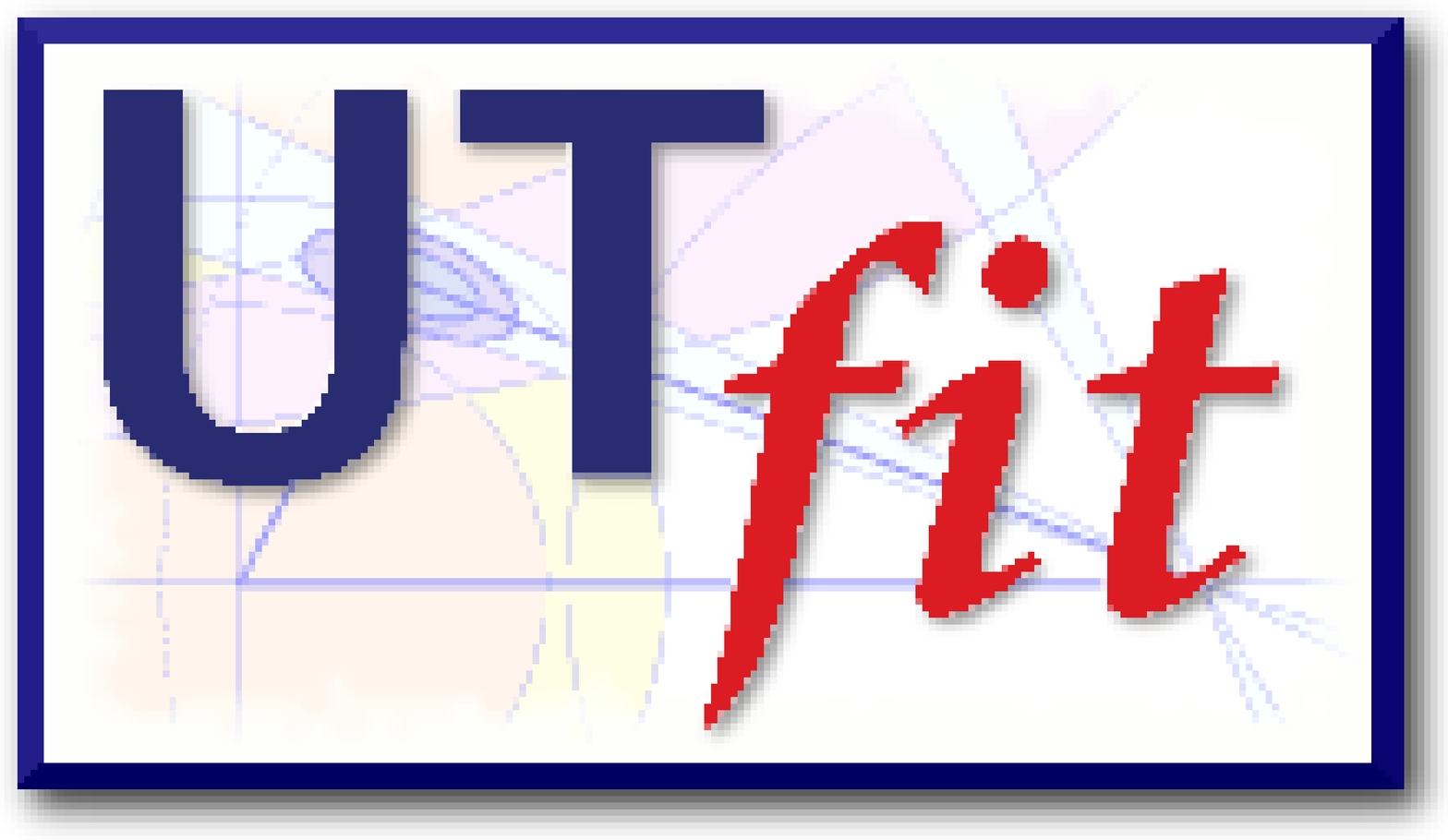}
 \end{center}
\end{figure}
\vspace*{-1cm}
}

\collaboration{\textbf{UT}\textit{fit} Collaboration}
\noaffiliation
\author{M.~Bona}
\affiliation{Laboratoire d'Annecy-le-Vieux de Physique des Particules
LAPP, IN2P3/CNRS, Universit{\'e} de Savoie, BP 110 F-74941
Annecy-le-Vieux Cedex, France} 
\author{M.~Ciuchini}
\affiliation{Dip.~di Fisica, Universit\`a di Roma Tre
      and INFN,  Sez.~di Roma Tre, I-00146 Roma, Italy}
\author{E.~Franco}
\affiliation{Dip.~di Fisica, Universit\`a di Roma ``La Sapienza'' and
  INFN, Sez.~di Roma, I-00185 Roma, Italy}
\author{V.~Lubicz}
\affiliation{Dip.~di Fisica, Universit\`a di Roma Tre
      and INFN,  Sez.~di Roma Tre, I-00146 Roma, Italy}
\author{G.~Martinelli}
\affiliation{Dip.~di Fisica, Universit\`a di Roma ``La Sapienza'' and
  INFN, Sez.~di Roma, I-00185 Roma, Italy}
\author{F.~Parodi}
\affiliation{ Dip.~di Fisica, Universit\`a di Genova and INFN, I-16146
  Genova, Italy} 
\author{M.~Pierini}
\affiliation{CERN, CH-1211 Geneva 23, Switzerland}
\author{P.~Roudeau}
\affiliation{Laboratoire de l'Acc\'el\'erateur Lin\'eaire, IN2P3-CNRS et
  Univ.~de Paris-Sud, BP 34, 
      F-91898 Orsay Cedex, France}
\author{C.~Schiavi}
\affiliation{ Dip.~di Fisica, Universit\`a di Genova and INFN, I-16146
  Genova, Italy} 
\author{L.~Silvestrini}
\affiliation{Dip.~di Fisica, Universit\`a di Roma ``La Sapienza'' and
  INFN, Sez.~di Roma, I-00185 Roma, Italy}
\author{V.~Sordini}
\affiliation{Laboratoire de l'Acc\'el\'erateur Lin\'eaire, IN2P3-CNRS et
  Univ.~de Paris-Sud, BP 34, 
      F-91898 Orsay Cedex, France}
\author{A.~Stocchi}
\affiliation{Laboratoire de l'Acc\'el\'erateur Lin\'eaire, IN2P3-CNRS et
  Univ.~de Paris-Sud, BP 34, 
      F-91898 Orsay Cedex, France}
\author{V.~Vagnoni}
\affiliation{INFN, Sez.~di Bologna,  I-40126 Bologna, Italy}

\begin{abstract}
  We update the constraints on new-physics contributions to $\Delta
  F=2$ processes from the generalized unitarity triangle analysis,
  including the most recent experimental developments. Based on these
  constraints, we derive upper bounds on the coefficients of the most
  general $\Delta F=2$ effective Hamiltonian. These upper bounds can
  be translated into lower bounds on the scale of new physics that
  contributes to these low-energy effective interactions. We point out
  that, due to the enhancement in the renormalization group evolution
  and in the matrix elements, the coefficients of non-standard
  operators are much more constrained than the coefficient of the
  operator present in the Standard Model.  Therefore, the scale of new
  physics in models that generate new $\Delta F=2$ operators, such as
  next-to-minimal flavour violation, has to be much higher than the
  scale of minimal flavour violation, and it most probably lies beyond
  the reach of direct searches at the LHC.
\end{abstract}

\maketitle

\section{Introduction}
\label{sec:intro}

Starting from the pioneering measurements of the $K^0 - \bar K^0$ mass
difference $\Delta m_K$ and of the CP-violating parameter
$\varepsilon_K$, continuing with the precision measurements of the $B_d
- \bar B_d$ mixing parameters $\Delta m_{B_d}$ and $\sin 2 \beta$ and with
the recent determination of the $B_s - \bar B_s$ oscillation frequency
$\Delta m_{B_s}$ and the first bounds on the mixing phase $- 2 \beta_s$, until
the very recent evidence of $D^0 - \bar D^0$ mixing, $\Delta F=2$
processes have always provided some of the most stringent constraints
on New Physics (NP).

For example, it has been known for more than a quarter of century that
supersymmetric extensions of the Standard Model (SM) with generic
flavour structures are strongly constrained by $K^0-\bar K^0$ mixing and
CP violation \cite{susyjurassic}. The constraints from $K^0-\bar K^0$
mixing are particularly stringent for models that generate transitions
between quarks of different chiralities
\cite{LRjurassic,pellicanino,pellicani}.  More recently, it has been
shown that another source of enhancement of chirality-breaking
transitions lies in the QCD corrections \cite{bagger}, now known at
the Next-to-Leading Order (NLO) \cite{NLOgen,BBSUSY}.

Previous phenomenological analyses of $\Delta F=2$ processes in
supersymmetry \cite{SUSYKKbar,SUSYBBbar} were affected by a large
uncertainty due to the SM contribution, since no determination of the
Cabibbo-Kobayashi-Maskawa~\cite{ckm} (CKM) CP-violating phase was
available in the presence of NP. A breakthrough was possible with the
advent of $B$ factories and the measurement of time-dependent CP
asymmetries in $B$ decays, allowing for a simultaneous determination
of the CKM parameters and of the NP contributions to $\Delta F=2$
processes in the $K^0$ and $B_d$ sectors
\cite{UTfitNP05,UTfitNP06,othernp}. Furthermore, the
Tevatron experiments have provided the first measurement of $\Delta
m_{B_s}$ and the first bounds on the phase of $B_s - \bar B_s$ mixing.
Combining all these ingredients, we can now determine allowed ranges
for all NP $\Delta F=2$ amplitudes in the down-quark sector.

To complete the picture, the recent evidence of $D^0-\bar D^0$ mixing
allows to constrain NP contributions to the $\Delta C=2$
amplitude~\cite{noiddbar,altriddbar}.

Our aim in this work is to consider the most general effective
Hamiltonian for $\Delta F=2$ processes
($\mathcal{H}_\mathrm{eff}^{\Delta F=2}$) and to translate the
experimental constraints into allowed ranges for the Wilson
coefficients of $\mathcal{H}_\mathrm{eff}^{\Delta F=2}$. These
coefficients in general have the form
\begin{equation}
  \label{eq:cgenstruct}
  C_i (\Lambda) = \frac{F_i L_i}{\Lambda^2}\,
\end{equation}
where $F_i$ is a function of the (complex) NP flavour couplings, $L_i$
is a loop factor that is present in models with no tree-level Flavour
Changing Neutral Currents (FCNC), and $\Lambda$ is the scale of NP,
\emph{i.e.}  the typical mass of the new particles mediating
$\Delta F=2$ transitions. For a generic strongly-interacting theory
with arbitrary flavour structure, one expects $F_i \sim L_i \sim 1$ so
that the allowed range for each of the $C_i(\Lambda)$ can be
immediately translated into a lower bound on $\Lambda$. Specific
assumptions on the flavour structure of NP, for example Minimal
\cite{mfv,uut,dambrosio} or Next-to-Minimal \cite{papucci} Flavour
Violation (MFV or NMFV), correspond to particular choices of the $F_i$
functions, as detailed below.

Our study is analogous to the operator analysis of electroweak
precision observables \cite{barbierietal}, but it
provides much more stringent bounds on models with non-minimal flavour
violation. In particular, we find that the scale of heavy particles
mediating tree-level FCNC in models of NMFV must lie above $\sim 60$ TeV,
making them undetectable at the LHC.  This bound applies for instance
to the Kaluza-Klein excitations of gauge bosons in a large class of
models with (warped) extra dimensions \cite{soni}. Flavour physics
remains the main avenue to probe such extensions of the SM.

The paper is organised as follows. In Sec.~\ref{sec:exp} we briefly
discuss the experimental novelties considered in our analysis. In
Sec.~\ref{sec:mi} we present updated results for the analysis of the
Unitarity Triangle (UT) in the presence of NP, including the
model-independent constraints on $\Delta F=2$ processes, following
closely our previous analyses \cite{UTfitNP05,UTfitNP06}. In
Sec.~\ref{sec:EH} we discuss the structure of
$\mathcal{H}_\mathrm{eff}^{\Delta F=2}$, the definition of the models
we consider and the method used to constrain the Wilson
coefficients. In Sec.~\ref{sec:results} we present our results for the
Wilson coefficients and for the scale of NP. Conclusions are drawn in
Sec.~\ref{sec:concl}.

\section{Experimental input}
\label{sec:exp}

We use the same experimental input as Ref.~\cite{UTfitNP06}, updated
after the Winter $'07$ conferences. We collect all the numbers used
throughout this paper in Tables~\ref{tab:expinput} and
\ref{tab:hadinput}. We include the following novelties: the most
recent result for $\Delta m_s$~\cite{dmsCDF}, the semileptonic
asymmetry in $B_s$ decays $A_\mathrm{SL}^{s}$~\cite{ASLD0} and the dimuon
charge asymmetry $A_\mathrm{SL}^{\mu\mu}$ from D$\O$~\cite{ACHD0} and CDF~\cite{ASLCDF},
the measurement of the $B_s$ lifetime from
flavour-specific final states~\cite{tauBsflavspec}, the determination
of $\Delta \Gamma_s /\Gamma_s$ from the time-integrated angular
analysis of $B_s \to J/\psi \phi$ decays by CDF~\cite{DGoGCDF}, the
three-dimensional constraint on $\Gamma_s$, $\Delta \Gamma_s$, and the
phase $\phi_s$ of the $B_s$--$\bar B_s$ mixing amplitude from the
time-dependent angular analysis of $B_s\to J/\psi \phi$ decays by
D$\O$~\cite{DGoGD0}.

\begin{table}[htb!]
\begin{center}
\begin{tabular}{lccc}
\hline\hline

Parameter  &  ~~~~~~Value~~~~~~  & ~~~~~~Gaussian ($\sigma$)~~~~~~      &   Uniform     \\
          &         &                          & (half-width)   \\ 
\hline\hline
$\lambda$  &  0.2258 &  0.0014                  &    -          \\ 
$\left |V_{cb} \right | \times 10^3$(excl.) & $ 39.1$            
     & $0.6$     & $ 1.7$          \\
$\left |V_{cb} \right | \times 10^3$(incl.) & $ 41.7 $            
     & $0.7$     & -  \\
$\left |V_{ub} \right |\times 10^4$(excl.)             & $ 34$           
     & $4$     & -  \\
$\left |V_{ub} \right |\times 10^4$(incl.)        & $ 43.1 $           
     & $3.9$     &        -              \\ 
$\Delta m_d$ (ps$^{-1}$)                  & $0.507$            
     & $0.005$   &        -              \\
$\Delta m_s$   (ps$^{-1}$)   &  $17.77$            
     & $0.12$   &  - \\
$\epsilon_K\times 10^3$                                 & $2.232$
     & $0.007$   &          -           \\
$\sin 2 \beta$     &
\multicolumn{3}{l}{see Winter '07 analysis at \texttt{http://www.utfit.org} }\\
$\cos 2 \beta$         &
\multicolumn{3}{l}{see Winter '07 analysis at \texttt{http://www.utfit.org}}\\
$\alpha$         &
\multicolumn{3}{l}{see Winter '07 analysis at \texttt{http://www.utfit.org}} \\
$\gamma$         & \multicolumn{3}{l}{see Winter '07 analysis at \texttt{http://www.utfit.org}} \\
$2 \beta + \gamma$& \multicolumn{3}{l}{see Winter '07 analysis at \texttt{http://www.utfit.org}} \\
$\overline m_t$ (GeV)  & $161.2$ & $1.7$ &          -            \\
$\alpha_s(M_Z)$                                  & 0.119       
     & 0.003                    &          -         \\
$\tau_{B_s}$   (ps)  & 1.39 &   0.12 &   - \\
$\tau_{B_s}^{FS}$   (ps)  & 1.454 &   0.040 &   - \\
$A_\mathrm{SL}^d$ &  -0.0005 & 0.0056 &  - \\
$A_\mathrm{SL}^s$ &  0.0245 & 0.0196 &  - \\
$A_\mathrm{SL}^{\mu\mu}$ & -0.0043 & 0.0030 &  - \\
$\Delta \Gamma_d/\Gamma_d$ & 0.009 & 0.037 & - \\
$\Delta \Gamma_s/\Gamma_s$ & 0.65 & 0.33 & - \\ \hline
$\phi_s$ [rad] & -0.79 & 0.56 & $\phi_s$-$ \tau_{B_s}$ corr. 0.727 \\
$ \tau_{B_s}$   (ps)  & 1.49 & 0.08 & $\tau_{B_s}$-$\Delta \Gamma_s/\Gamma_s$ corr. -0.172  \\
$\Delta \Gamma_s/\Gamma_s$ & 0.17 &  0.09 & $\Delta
\Gamma_s/\Gamma_s$-$\phi_s$ corr. -0.188\\
\hline\hline
\end{tabular} 
\end{center}
\caption{Values of the experimental input used in our analysis.
  The Gaussian and the flat contributions to the
  uncertainty are given in the third and fourth columns 
  respectively (for details on the statistical treatment
  see~\cite{noi}). See text
  for details.} 
\label{tab:expinput} 
\end{table}

\begin{table}[htbp!]
\begin{center}
\begin{tabular}{@{}lccc}
\hline\hline
Parameter  &  Value  & Gaussian ($\sigma$)      &   Uniform     \\
          &         &                          & (half-width)   \\ 
\hline\hline
$F_{D}$ (MeV)            & $201$                         
     & $3$                 &          $17$            \\ 
$F_{B_s}\sqrt{B_s}$ (MeV)            & $262$                         
     & $35$                 &          -            \\ 
$\xi=\frac{F_{B_s}\sqrt{B_s}}{F_{B_d}\sqrt{B_d}}$             & 1.23        
          & 0.06                     &  -          \\
$\hat B_K$                                  & 0.79                              
     & 0.04                     &     0.08              \\
$\overline m_b$ (GeV)      & 4.21    & 0.08          &          -            \\
$\overline m_c$ (GeV)    & 1.3      & 0.1     &          -            \\ 
$R_1$ & 1 & - & - \\
$R_2$ & -12.9 & 3.0 & - \\
$R_3$ & 3.98 & 0.89 & - \\
$R_4$ & 20.8 & 4.4 & - \\
$R_5$ & 5.2 & 1.2 & - \\
$B_1^D$ & 0.865 & 0.02 & 0.015 \\
$B_2^D$ & 0.82 &  0.03 & 0.01 \\
$B_3^D$ & 1.07 &  0.05 & 0.08 \\
$B_4^D$ & 1.08 &  0.02 & 0.02 \\
$B_5^D$ & 1.455 & 0.03 & 0.075 \\
$B_1^B$ & 0.88 & 0.04 & 0.10 \\
$B_2^B$ & 0.82 & 0.03 & 0.09 \\
$B_3^B$ & 1.02 & 0.06 & 0.13 \\
$B_4^B$ & 1.15 & 0.03 & 0.13 \\
$B_5^B$ & 1.99 & 0.04 & 0.24 \\
\hline\hline
\end{tabular} 
\end{center}
\caption{Values of the hadronic parameters used in our analysis.
The Gaussian and the flat contributions to the
uncertainty are given in the third and fourth columns 
respectively (for details on the statistical treatment
see~\cite{noi}). See the text for details.}
\label{tab:hadinput} 
\end{table}

The use of $\Delta \Gamma_s /\Gamma_s$, from the time-integrated
angular analysis of $B_s \to J/\psi \phi$ decays, is described in
Ref.~\cite{UTfitNP06}. In this paper, we only use the CDF measurement
as input, since the D$\O$ analysis is now superseded by the new
time-dependent study.  The latter provides the first direct constraint
on the $B_s$--$\bar B_s$ mixing phase, but also a simultaneous bound
on $\Delta \Gamma_s$ and $\Gamma_s$. We implemented the full $3 \times
3$ correlation matrix. The time-dependent analysis determines the
$B_s$--$\bar B_s$ mixing phase with a four-fold
ambiguity.~\footnote{Notice that the definition used by D$\O$ is the
  one of Ref.~\cite{dighe}, namely $\phi_s=2\beta_s=2 \arg
  (-(V_{ts}V_{tb}^*)/(V_{cs}V_{cb}^*)))$ in the SM. Notice also that
  in the arXiv version of Ref.~\cite{DGoGD0} the definition of
  $\phi_s$ is unclear.} First of all, the $B_s$ mesons are untagged,
so the analysis is not directly sensitive to $\sin \phi_s$, resulting
in the ambiguity $(\phi_s,\cos \delta_{1,2}) \leftrightarrow
(-\phi_s,-\cos \delta_{1,2})$, where $\delta_{1,2}$ represent the
strong phase differences between the transverse polarization and the
other ones. Second, at fixed sign of $\cos \delta_{1,2}$, there is the
ambiguity $(\phi_s,\Delta \Gamma_s) \leftrightarrow
(\phi_s+\pi,-\Delta \Gamma_s)$. Concerning the strong phases
$\delta_i$, there is a two-fold ambiguity corresponding to $\delta_i
\to \pi - \delta_i$. The two experimental determinations are roughly
$\delta_2 \sim 0$, $\delta_1 \sim \pi$ and $\delta_1 \sim 0$,
$\delta_2 \sim \pi$. In the literature it is often found that
factorization corresponds to the first choice
\cite{dighe,digheold,nierste}. However, we find that factorization
predicts $\delta_1 \sim 0$, $\delta_2 \sim \pi$
\cite{kramer,neubert,beneke}. This result is also compatible with the
BaBar measurement in $B \to J/\Psi K^*$ \cite{babarjpsikst}, which can
be related to $B_s \to J/\Psi \phi$ using $SU(3)$ and neglecting
singlet contributions.\footnote{In the first version of this
  manuscript, we stated that factorization disagreed with $SU(3)$, based
  on the factorization prediction in
  Refs.~\cite{dighe,digheold,nierste}.} However, waiting for future,
more sophisticated experimental analyses which could resolve this
ambiguity, we prefer to be conservative and keep the four-fold
ambiguity in our analysis.

The use of $\Delta m_s$ was already discussed in
Ref.~\cite{UTfitNP06}.  The only difference with respect to that is
the update of the experimental inputs: we now use the improved
measurement by CDF~\cite{dmsCDF}, and we take $\tau_{B_s}$ only from the study
of $B_s$ decays to CP eigenstates~\cite{tauBsCP}.
The value of $\tau_{B_s}$ obtained from $B_s$ decaying to
flavour-specific final states, using a single exponential in the fit,
is related to the values of $\Gamma_s$ and $\Delta \Gamma_s$ by the
relation~\cite{teoTauBs}
\begin{equation}
  \tau_{B_s}^{FS} = \frac{1}{\Gamma_s}\frac{1+\left(\frac{\Delta
        \Gamma_s}{2\Gamma_s}\right
    )^2}{1-\left(\frac{\Delta\Gamma_s}{2\Gamma_s}\right )^2}\,,
\end{equation}
which provides an independent constraint on $\Delta\Gamma_s/\Gamma_s$.
We compute $\Delta \Gamma_s$ and $A_\mathrm{SL}^{s}$ using eq.~(7) of
Ref.~\cite{UTfitNP06} (recalling that $A_\mathrm{SL}^s=2 (1-\vert
q/p\vert_s)$. Following Ref.~\cite{nir}, we use the value of
$A_\mathrm{SL}^{\mu\mu}$ recently
presented by D$\O$~\cite{ACHD0} and CDF~\cite{ASLCDF}, in the form
\begin{equation}
  \label{eq:ach}
  A_\mathrm{SL}^{\mu\mu}=
  \frac{f_d \chi_{d0} A_\mathrm{SL}^d+f_s \chi_{s0} A_\mathrm{SL}^s}
       {f_d \chi_{d0}+f_s \chi_{s0}}\,,
\end{equation}
with $f_d = 0.397 \pm 0.010$, $f_s = 0.107 \pm
0.011$, $\chi_{q0}= (\chi_q+\bar \chi_q)/2$.  $\chi_{q}$ and $\bar
\chi_{q}$ are computed using equations (3)-(5) of
Ref.~\cite{UTfitNP06}.\footnote{To combine the CDF and D$\O$ measurements,
we have converted the value for $A$ defined in Ref.~\cite{ACHD0} into a value
for $A_\mathrm{SL}^{\mu\mu}$.}

Finally, concerning $D^0 -\bar D^0$ mixing, we use as input the
results for the NP amplitude obtained in Ref.~\cite{noiddbar}
combining the experimental information from Refs.~\cite{ddbarexp}.

\section{UT analysis and constraints on NP}
\label{sec:mi}

The contribution of NP to $\Delta F=2$ transitions
can be parameterized in a model-independent way as the ratio of the
full (SM+NP) amplitude to the SM one. In this way, we can
define the parameters $C_{B_q}$ and $\phi_{B_q}$ ($q = d,\,
s$) as \cite{cfactors}:
\begin{equation} 
  C_{B_q} \, e^{2 i \phi_{B_q}} =\frac{\langle
    B_q|H_\mathrm{eff}^\mathrm{full}|\bar{B}_q\rangle} {\langle
    B_q|H_\mathrm{eff}^\mathrm{SM}|\bar{B}_q\rangle}
\,, 
  \label{eq:paranp}
\end{equation}
and write all the measured observables as a function of these
parameters and the SM ones ($\bar\rho$, $\bar\eta$, and additional
parameters such as masses, form factors, and decay constants). Details
are given in Refs.~\cite{UTfitNP05,UTfitNP06}. In a similar way, one
can write
\begin{equation}
  C_{\epsilon_K} = \frac{\mathrm{Im}[\langle
    K^0|H_{\mathrm{eff}}^{\mathrm{full}}|\bar{K}^0\rangle]}
  {\mathrm{Im}[\langle
    K^0|H_{\mathrm{eff}}^{\mathrm{SM}}|\bar{K}^0\rangle]}\,,\qquad
  C_{\Delta m_K} = \frac{\mathrm{Re}[\langle
    K^0|H_{\mathrm{eff}}^{\mathrm{full}}|\bar{K}^0\rangle]}
  {\mathrm{Re}[\langle
    K^0|H_{\mathrm{eff}}^{\mathrm{SM}}|\bar{K}^0\rangle]}\,.
  \label{eq:ceps}
\end{equation}
Concerning $\Delta m_K$, to be conservative, we add to the
short-distance contribution a possible long-distance one that varies
with a uniform distribution between zero and the experimental value of
$\Delta m_K$.
\begin{figure}[tb!]
\begin{center}
\includegraphics[width=0.65\textwidth]{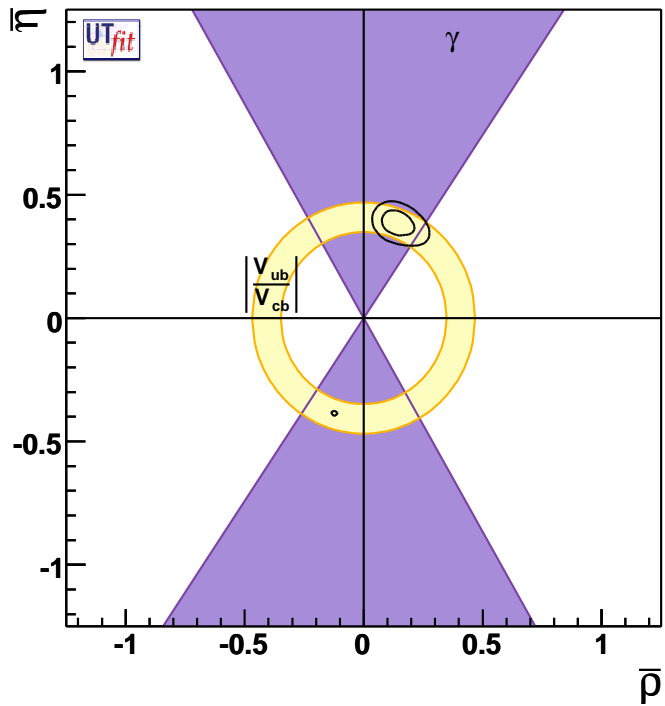}
\caption{%
  Determination of $\bar\rho$ and $\bar \eta$ from the NP generalized
  analysis. $68\%$ and $95\%$ probability regions for  $\bar\rho$ and
  $\bar \eta$ are shown, together with the $2\,\sigma$ contours given
  by the tree-level determination of $\vert V_{ub}\vert$ and $\gamma$.}
\label{fig:NPrhoeta}
\end{center}
\end{figure}

\begin{table}[tb!]
\begin{center}
\begin{tabular}{c|c|c}
\hline\hline
    Parameter               &   $68 \%$ Probability  &   $95 \%$
    Probability    \\
\hline \hline
$C_{B_d}$ & $1.05 \pm 0.34$ & $[0.53,2.05]$ \\
$\phi_{B_d} [^{\circ}]$ & $-3.4 \pm 2.2$  & $[-8.3,1.6]$ \\
$C_{B_s}$ & $1.11 \pm 0.32$ & $[0.63,2.07]$ \\
$\phi_{B_s} [^{\circ}]$ &   $(-69 \pm 14)$ $\cup$ $(-20 \pm 14)$ $\cup$ $(20
\pm 5)$ $\cup$ $(72 \pm 8)$
& $[-86,-46]$ $\cup$ $[-43,35]$ $\cup$ $[56,87]$ \\
$C_{\Delta m_K}$ & $0.93 \pm 0.32$ & $[0.51,2.07]$ \\
$C_{\epsilon_K}$ & $0.92 \pm 0.14$ & $[0.66,1.31]$ \\
\hline \hline
$\overline {\rho}$  &  $0.140\pm0.046$  & $[0.049,0.258]$ \\
$\overline {\eta}$  &  $0.384\pm0.035$  & $[0.304,0.460]$ \\
$\alpha [^{\circ}]$ &  $86\pm 7$ &  $[73,104]$ \\
$\beta [^{\circ}]$ &  $24.3\pm 2.0$ &  $[19.9,28.6]$\\
$\gamma[^{\circ}]$ &  $70\pm 7$ &  $[53,83]$\\
$\rm{Re} {\lambda}_t$[$10^{-5}$] &  $-31.3\pm 1.8$ &  $[-35,-27]$\\
$\rm{Im} {\lambda}_t$[$10^{-5}$] &  $14.8\pm 1.3$ &  $[12.3,17.3]$\\
$\vert V_{ub}\vert[10^{-3}]$  & $3.91\pm 0.28$ &  $[3.36,4.44]$\\
$\vert V_{cb}\vert[10^{-2}]$  & $4.09\pm 0.05$ &  $[3.98,4.18]$ \\
$\vert V_{td}\vert[10^{-3}]$  & $8.7\pm 0.5$ & $[7.6,9.6]$ \\
$|V_{td}/V_{ts}|$  & $0.217\pm 0.012$ & $[0.187,0.240]$ \\
$R_b$  & $0.413\pm 0.030$ & $[0.354,0.471]$ \\
$R_t$  & $0.943\pm 0.048$ & $[0.819,1.037]$\\
$\sin2\beta$ & $0.748 \pm 0.044$ & $[0.643,0.841]$\\
$\sin2\beta_s$ & $0.0409 \pm 0.0038$ & $[0.0321,0.0494]$\\
\hline\hline
\end{tabular}
\end{center}
\caption {Determination of NP and UT parameters
from the UT fit.}
\label{tab:NP}
\end{table} 

We perform a global analysis using the method of Ref.~\cite{noi} and
determine simultaneously $\bar\rho$, $\bar\eta$, $C_{B_q}$,
$\phi_{B_q}$, $C_{\epsilon_K}$ and $C_{\Delta m_K}$ using flat
a-priori distributions for these parameters. The resulting probability
density function (p.d.f.) in the $\bar\rho-\bar\eta$ plane is shown in
Fig.~\ref{fig:NPrhoeta}. Only a small region close to the result of
the SM fit survives. The mirror solution in the third quadrant is
suppressed down to about $5\%$ probability by the measurements of
$A_\mathrm{SL}^d$ and $A_\mathrm{SL}^{\mu\mu}$. The results for
$\bar\rho$ and $\bar\eta$ reported in Tab.~\ref{tab:NP} are at a level
of accuracy comparable to the SM fit \cite{UTfitSM06}, so that the SM
contribution to FCNC processes \emph{in the presence of arbitrary NP}
is bound to lie very close to the results of the SM \emph{in the
  absence of NP}. This result represents a major improvement in the
study of FCNC processes beyond the SM, and opens up the possibility of
precision studies of flavour processes in the presence of NP.

\begin{figure}[tb!]
\begin{center}
\includegraphics[width=0.45\textwidth]{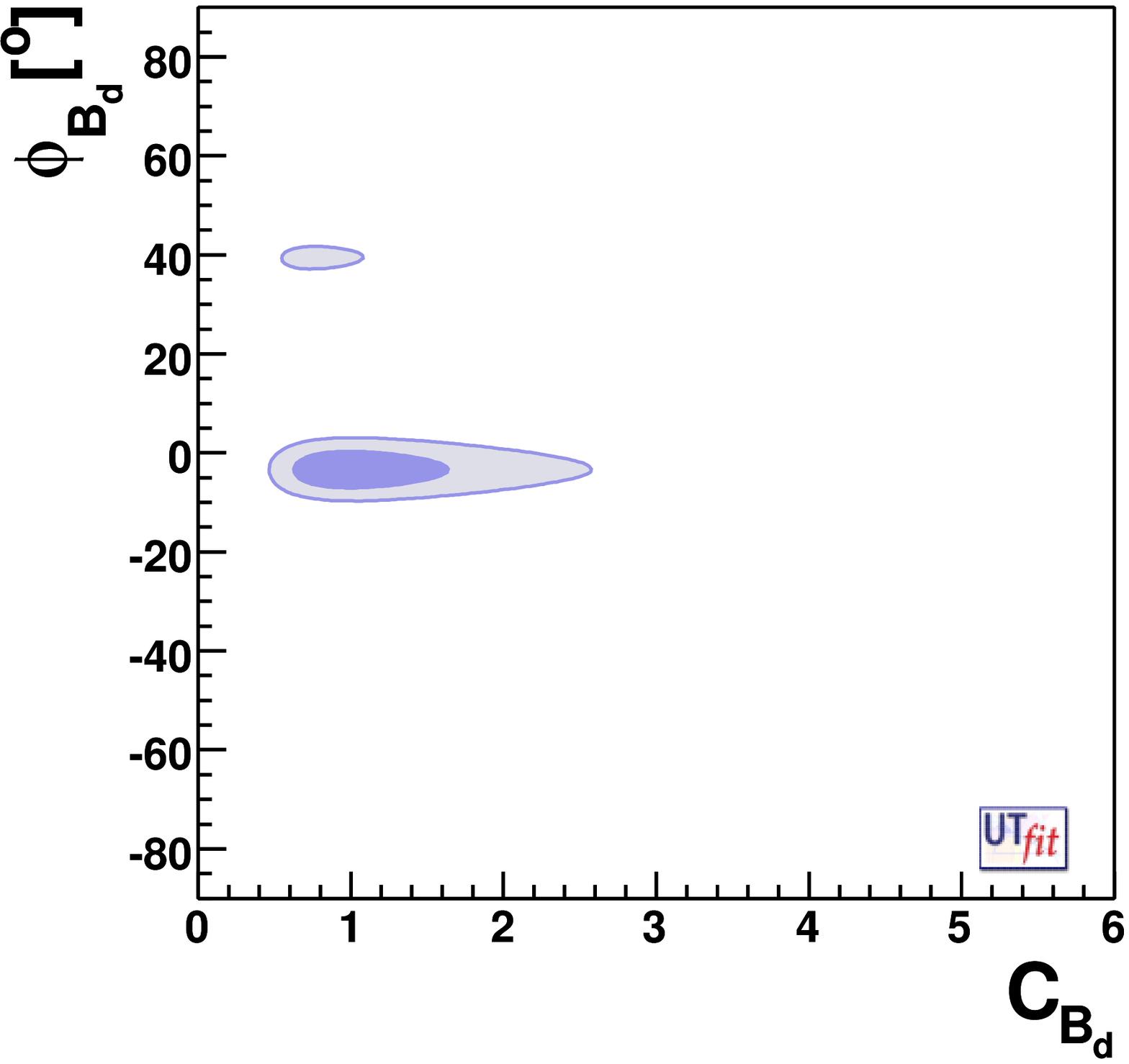}
\includegraphics[width=0.45\textwidth]{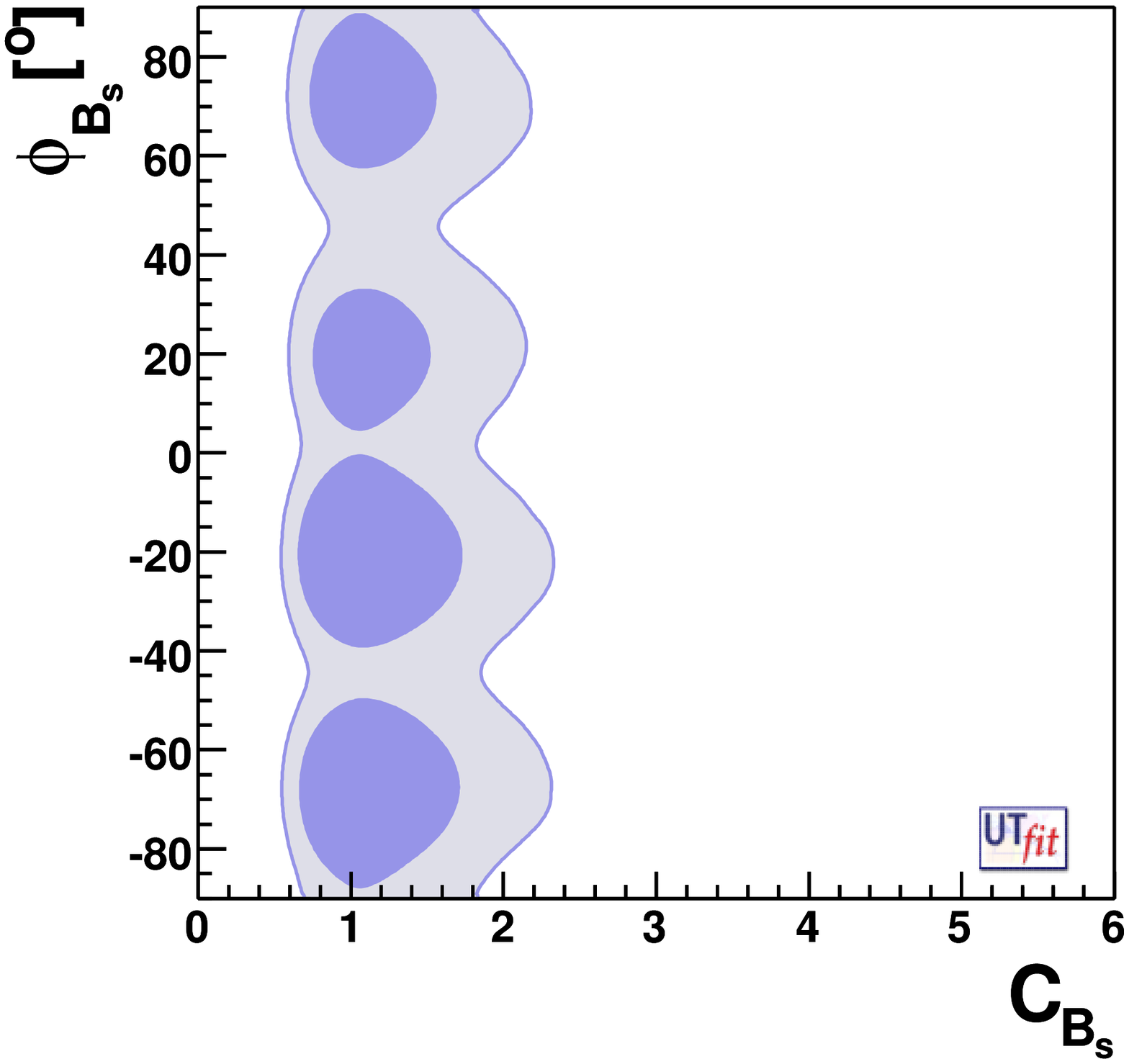}
\includegraphics[width=0.45\textwidth]{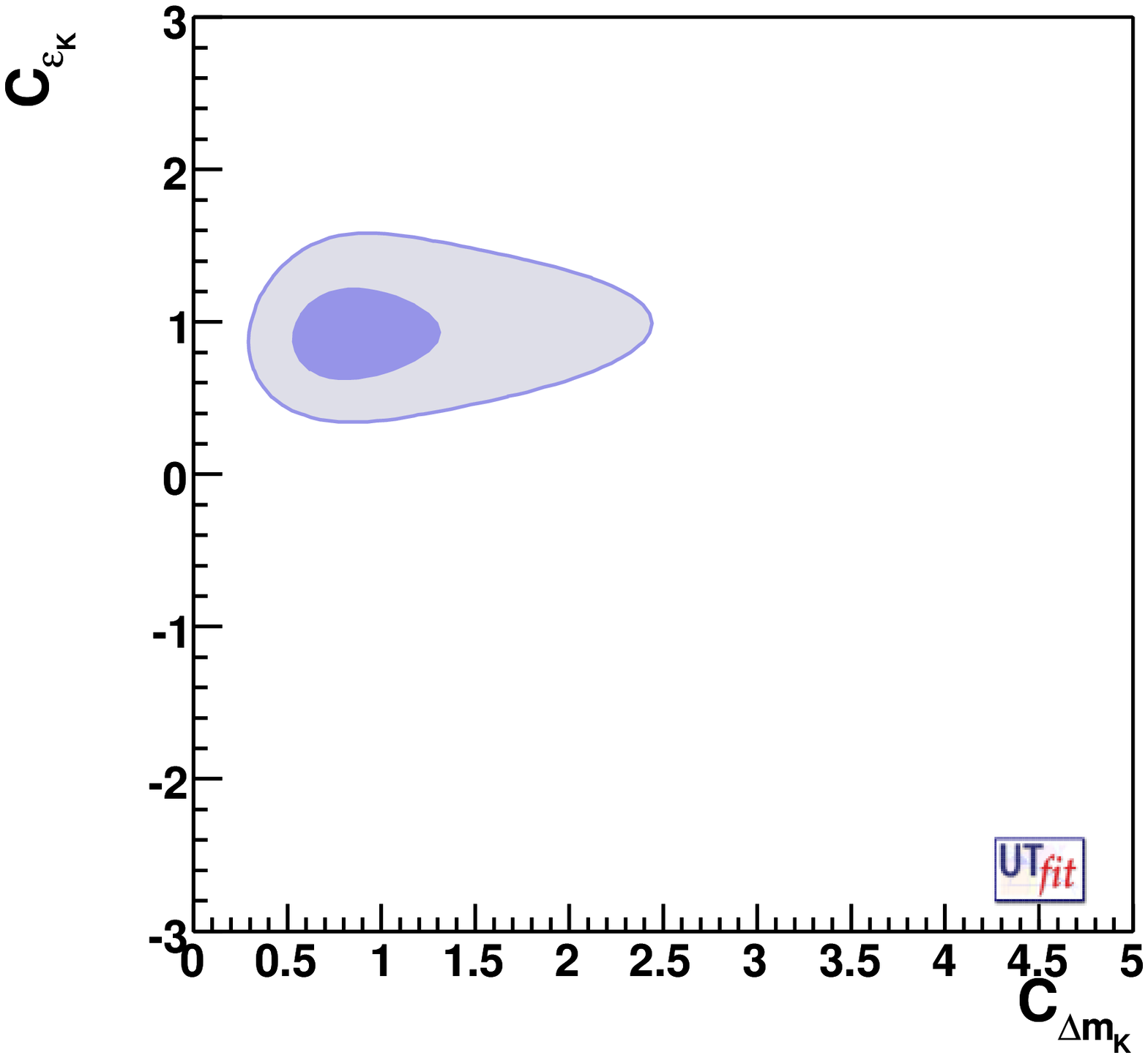}
\caption{%
  Constraints on $\phi_{B_d}$ vs. $C_{B_d}$,
  $\phi_{B_s}$ vs. $C_{B_s}$ and $C_{\epsilon_K}$ vs $C_{\Delta m_K}$
  from the NP generalized analysis.}
\label{fig:NP}
\end{center}
\end{figure}

The constraining power of this analysis is evident in the results for
the NP parameters given in Tab.~\ref{tab:NP} and shown in
Fig.~\ref{fig:NP}. Compared to our previous analysis in
Ref.~\cite{UTfitNP06}, and to similar analyses in the literature
\cite{otherBs,nir}, we see that the additional experimental input
discussed above improves considerably the determination of the phase
of the $B_s - \bar B_s$ mixing amplitude. The fourfold ambiguity
inherent in the untagged analysis of Ref.~\cite{DGoGD0} is somewhat
reduced by the measurements of $A_\mathrm{SL}^s$ and
$A_\mathrm{SL}^{\mu\mu}$, which prefer negative values of
$\phi_{B_s}$.\footnote{With respect to Ref.~\cite{nierste}, we find
  that the inclusion of $A_\mathrm{SL}^{\mu\mu}$ has a weaker impact
  in reducing the ambiguity coming from several small differences in
  the analysis (theoretical assumptions on NP in $A_{SL}^d$, presence
  of NP in penguin contributions to $A_{SL}^{d,s}$, inclusion of the
  CDF measurement of $A_\mathrm{SL}^{\mu\mu}$, etc.).}
Ref.~\cite{nierste} recently claimed a $2\sigma$ deviation from zero
in $\phi_{B_s}$, taking the sign of $\cos \delta_{1,2}$ from
factorization. We confirm that, with the same assumptions of
Ref.~\cite{nierste} on strong phases,\footnote{We find that
  factorization gives $\delta_1 \sim 0$ and $\delta_2 \sim \pi$,
  resolving the ambiguity of the $D0$ untagged analysis in favour of
  $\phi_s \sim 0.79$ for positive $\Delta \Gamma$, while
  Ref.~\cite{nierste} uses $\delta_1 \sim \pi$, $\delta_2 \sim 0$ and
  $\phi_s \sim -0.79$. However, this sign difference in $\phi_s$ is
  compensated by the fact that $\phi_s$ as defined in 
  Ref.~\cite{nierste} should be compared to $-\phi_s$ as measured by
  $D0$.} the deviation from zero of $\phi_{B_s}$ slightly exceeds
$2\sigma$. Without assuming strong phases from factorization, we find
a deviation of $\phi_{B_s}$ from zero of $1.4~\sigma$, of the same
size of the deviation found experimentally by Ref.~\cite{DGoGD0}.

It is important to stress that the information contained in these
constraints does not rely on any specific model for NP. The list of
applications in NP phenomenology is rich. For instance, restricting to
the case of SUSY models, the two-dimensional constraint of the $b \to
s$ ($b \to d$) sector can be translated into a limit on the
mass-insertion complex parameters $\delta^d_{23}$ ($\delta^d_{13}$),
using the NLO supersymmetric expression of the $B_q$--$\bar B_q$
mixing amplitude~\cite{BBSUSY}. This bound, combined with the constraint
from $b \to s$ decays~\cite{massinsertion}, allows to obtain the best
available information on the off-diagonal terms of the squark mass
matrix~\cite{valentina}. However, in the following, rather than
considering an explicit model for NP, we perform a general analysis
based on the most general Hamiltonian for $\Delta F=2$ processes.

\section{NP contributions to $\mathbf{\Delta F=2}$ processes}
\label{sec:EH}

The most general effective Hamiltonians for $\Delta F=2$ processes
beyond the SM have the following form:
\begin{eqnarray}
  \label{eq:defheff}
  {\cal H}_{\rm eff}^{\Delta S=2}=\sum_{i=1}^{5} C_i\, Q_i^{sd} +
  \sum_{i=1}^{3} \tilde{C}_i\, \tilde{Q}_i^{sd} \\
  {\cal H}_{\rm eff}^{\Delta C=2}=\sum_{i=1}^{5} C_i\, Q_i^{cu} +
  \sum_{i=1}^{3} \tilde{C}_i\, \tilde{Q}_i^{cu} \nonumber \\
  {\cal H}_{\rm eff}^{\Delta B=2}=\sum_{i=1}^{5} C_i\, Q_i^{bq} +
  \sum_{i=1}^{3} \tilde{C}_i\, \tilde{Q}_i^{bq} \nonumber
\end{eqnarray}
where $q=d(s)$ for $B_{d(s)} - \bar B_{d(s)}$ mixing and 
\begin{eqnarray}
         \label{Qi}
         Q_1^{q_iq_j} & = & \bar q^{\alpha}_{jL} \gamma_\mu
         q^{\alpha}_{iL} \bar q^{\beta}_{jL} \gamma^\mu 
         q^{\beta}_{iL}\; ,
         \nonumber \\
         Q_2^{q_iq_j} & = & \bar q^{\alpha}_{jR}  q^{\alpha}_{iL} \bar
         q^{\beta}_{jR} q^{\beta}_{iL}\; , 
         \nonumber \\
         Q_3^{q_iq_j} & = & \bar q^{\alpha}_{jR}  q^{\beta}_{iL} \bar
         q^{\beta}_{jR} q^{\alpha}_{iL}\; , 
         \\
         Q_4^{q_iq_j} & = & \bar q^{\alpha}_{jR}  q^{\alpha}_{iL} \bar
         q^{\beta}_{jL} q^{\beta}_{iR}\; , 
         \nonumber \\
         Q_5^{q_iq_j} & = & \bar q^{\alpha}_{jR}  q^{\beta}_{iL} \bar
         q^{\beta}_{jL} q^{\alpha}_{iR}\; .
         \nonumber
\end{eqnarray}
Here $q_{R,L}=P_{R,L}\,q$, with $P_{R,L}=(1 \pm \gamma_5)/2$, and
$\alpha$ and $\beta$ are colour indices.  The operators
$\tilde{Q}_{1,2,3}^{q_iq_j}$ are obtained from the
$Q_{1,2,3}^{q_iq_j}$ by the exchange $ L \leftrightarrow R$. In the
following we only discuss the operators $Q_i$ as the results for
$Q_{1,2,3}$ apply to $\tilde{Q}_{1,2,3}$ as well.

The NLO anomalous dimension matrix has been computed in~\cite{NLOgen}.
We use the Regularisation-Independent anomalous dimension matrix in
the Landau gauge (also known as RI-MOM), since this scheme is used in
lattice QCD calculations of the matrix elements with non-perturbative
renormalization. 

The $C_i(\Lambda)$ are obtained by integrating out all new particles
simultaneously at the NP scale $\Lambda$.\footnote{Clearly, without
  knowing the masses of new particles, one cannot fix the scale
  $\Lambda$ of the matching. However, an iterative procedure quickly
  converges thanks to the very slow running of $\alpha_s$ at high
  scales.} We then have to evolve the
coefficients down to the hadronic scales $\mu_b=m_b=4.6$~GeV
($m_{b}\equiv m_{b}(\mu= m_{b})$ is the RI-MOM mass) for bottom
mesons, $\mu_D=2.8$~GeV for charmed mesons,
and $\mu_K=2$~GeV for Kaons, which are the renormalisation scales of
the operators used in lattice computations for the matrix elements
\cite{bpark,hep-lat/0110091}. 

We give here an analytic formula for the contribution to the $B_q -
\bar B_q$ mixing amplitudes induced by a given NP scale coefficient
$C_i(\Lambda)$, denoted by $\langle \bar B_q \vert {\cal H}_{\rm eff}
\vert B_q \rangle_i$, as a function of $\alpha_s(\Lambda)$:
\begin{equation}
\label{eq:magicbb}
\langle \bar B_q \vert {\cal H}_{\rm eff}^{\Delta B=2} \vert B_q \rangle_i = 
\sum_{j=1}^5 \sum_{r=1}^5
             \left(b^{(r,i)}_j + \eta \,c^{(r,i)}_j\right)
             \eta^{a_j} \,C_i(\Lambda)\, \langle \bar B_q \vert Q_r^{bq}
             \vert B_q \rangle\,,
\end{equation}
where $\eta=\alpha_s(\Lambda)/\alpha_s(m_t)$, the magic numbers $a_j$,
$b^{(r,i)}_j$ and $c^{(r,i)}_j$ and the matrix elements can be found
in eqs.~(10) and (12) of Ref.~\cite{SUSYBBbar} respectively. The
values of the $B^B_i$ parameters can be found in
Table~\ref{tab:hadinput}. A
similar formula holds for $D^0 - \bar D^0$ mixing, with the parameters
$B^D_i$ given in Table \ref{tab:hadinput} 
and the following magic numbers:
\begin{equation}
\label{eq:magic2}
\begin{array}{l l}
a_i=(0.286, -0.692, 0.787, -1.143, 0.143)& \\
& \\
b^{(11)}_i=(0.837, 0, 0, 0, 0),&
c^{(11)}_i=(-0.016,0,0,0,0),\\
b^{(22)}_i=(0,2.163,0.012,0,0),&
c^{(22)}_i=(0,-0.20,-0.002,0,0),\\
b^{(23)}_i=(0,-0.567,0.176,0,0),&
c^{(23)}_i=(0,-0.016,0.006,0,0),\\
b^{(32)}_i=(0,-0.032,0.031,0,0),&
c^{(32)}_i=(0,0.004,-0.010,0,0),\\
b^{(33)}_i=(0,0.008,0.474,0,0),&
c^{(33)}_i=(0,0.000,0.025,0,0),\\
b^{(44)}_i=(0,0,0,3.63,0),&
c^{(44)}_i=(0,0,0,-0.56,0.006),\\
b^{(45)}_i=(0,0,0,1.21,-0.19),&
c^{(45)}_i=(0,0,0,-0.29,-0.006),\\
b^{(54)}_i=(0,0,0,0.14,0),&
c^{(54)}_i=(0,0,0,-0.019,-0.016),\\
b^{(55)}_i=(0,0,0,0.045,0.839),&
c^{(55)}_i=(0,0,0,-0.009,0.018).\\
\end{array}
\end{equation}
All other magic numbers vanish.
Finally, for $K^0-\bar K^0$ mixing we obtain
\begin{equation}
  \label{eq:magickk}
  \langle \bar K^0 \vert {\cal H}_{\rm eff}^{\Delta S=2} \vert K^0 \rangle_i = 
  \sum_{j=1}^5 \sum_{r=1}^5
  \left(b^{(r,i)}_j + \eta \,c^{(r,i)}_j\right)
  \eta^{a_j} \,C_i(\Lambda)\, R_r \, \langle \bar K^0 \vert Q_1^{sd}
  \vert K^0 \rangle\,,
\end{equation}
where now the magic numbers can be found in eq.~(2.7) of
Ref.~\cite{SUSYKKbar}. We use the values in Table \ref{tab:hadinput} for
the ratios $R_r$ of the matrix elements of the NP operators $Q_r^{sd}$
over the SM one.  These values correspond to the average of the
results in Ref.~\cite{bpark}, applying a scaling factor to the errors
to take into account the spread of the available results.

To obtain the p.d.f.~for the Wilson coefficients at the NP scale
$\Lambda$, we switch on one coefficient at a time in each sector and
calculate its value from the result of the NP analysis presented in
sec.~\ref{sec:mi}.

As we discussed in eq.~(\ref{eq:cgenstruct}), the connection between
the $C_i(\Lambda)$ and the NP scale $\Lambda$ depends on the general
properties of the NP model, and in particular on the flavour structure
of the $F_i$. Assuming strongly interacting new particles, we have
from eq.~(\ref{eq:cgenstruct}) with $L_i=1$
\begin{equation}
  \label{eq:lambdagen}
  \Lambda=\sqrt{\frac{F_i}{C_i}}\,.
\end{equation}
Let us now discuss four notable examples:
\begin{itemize}
\item In the case of MFV with one Higgs doublet or two Higgs doublets
  with small or moderate $\tan \beta$, we have $F_1=F_\mathrm{SM}$ and
  $F_{i \neq 1}=0$, where $F_\mathrm{SM}$ is the combination of CKM
  matrix elements appearing in the top-quark mediated SM mixing
  amplitude, namely $(V_{tq} V_{tb}^*)^2$ for $B_q-\bar B_q$ mixing
  and $(V_{td} V_{ts}^*)^2$ for $\varepsilon_K$. $\Delta m_K$ and $D^0 -
  \bar D^0$ mixing do not give significant constraints in this scenario
  due to the presence of long-distance contributions. 
\item In the case of MFV at large $\tan \beta$, we have this additional
  contribution to $B_q - \bar B_q$ mixing \cite{dambrosio}:
  \begin{equation}
    \label{eq:mfvltb}
    C_4(\Lambda) = \frac{(a_0 + a_1) (a_0 + a_2)}{\Lambda^2} \lambda_b \lambda_q F_\mathrm{SM}\,,
  \end{equation}
  where $\lambda_{b,q}$ represent the corresponding Yukawa couplings,
  $a_{0,1,2}$ are $\tan \beta$-enhanced loop factors of
  $\mathcal{O}(1)$ and $\Lambda$ represents the NP scale corresponding
  to the non-standard Higgs bosons.
\item In the case of NMFV, we have $\vert F_i\vert = F_\mathrm{SM}$
  with an arbitrary phase~\cite{papucci} (following
  Ref.~\cite{dambrosio}, for $\Delta m_K$ and $D^0 - \bar D^0$ mixing we
  take $F_\mathrm{SM}=\vert V_{td} V_{ts}\vert^2$). This condition is
  realized in models in which right-handed currents also contribute to
  FCNC processes, but with the same hierarchical structure in the
  mixing angles as in the SM left-handed currents.  Given the
  order-of-magnitude equalities $m_d/m_b \sim \vert V_{td} \vert$,
  $m_s/m_b \sim \vert V_{ts} \vert$, bounds obtained in this scenario
  are also of interest for extra-dimensional models with FCNC
  couplings suppressed linearly with quark masses~\cite{soni}.
  Clearly, given the QCD and, for $K^0-\bar K^0$ mixing, chiral
  enhancement of NP operators, the constraints on the NP scale are
  much stronger for NMFV than for MFV, as shown explicitly in the next
  section.
\item For arbitrary NP flavour structures, we expect $\vert F_i \vert
  \sim 1$ with arbitrary phase. In this case, the constraints on the
  NP scale are much tighter due to the absence of the CKM suppression
  in the NP contributions.
\end{itemize}

\section{Results}
\label{sec:results}

In this Section, we present the results obtained for the four
scenarios described above. In deriving the lower bounds on the NP
scale $\Lambda$, we assume $L_i = 1$, corresponding to
strongly-interacting and/or tree-level NP. Two other interesting
possibilities are given by loop-mediated NP contributions proportional
to $\alpha_s^2$ or $\alpha_W^2$. The first case corresponds for
example to gluino exchange in the MSSM. The second case applies to all
models of SM-like loop-mediated weak interactions. To obtain the lower
bound on $\Lambda$ for loop-mediated contributions, one simply
multiplies the bounds we quote in the following by
$\alpha_s(\Lambda)\sim 0.1$ or by $\alpha_W \sim 0.03$.

Let us first consider MFV models and update our results presented in
Ref.~\cite{UTfitNP05,UTfitNP06}. In practice, the most convenient
strategy in this case is to fit the shift in the Inami-Lim top-quark
function entering $B_d$, $B_s$ and $K^0$ mixing. We fit for this shift
using the experimental measurements of $\Delta m_d$, $\Delta m_s$ and
$\epsilon_K$, after determining the parameters of the CKM matrix with
the universal unitarity triangle analysis \cite{uut}.\footnote{With
  respect to the original proposal of Ref.~\cite{uut}, we do not use
  the ratio $\Delta m_s/\Delta m_d$ in the fit in order to allow for
  Higgs-mediated contributions affecting $\Delta m_s$ at very large
  $\tan \beta$.} We obtain the following lower bounds at $95\%$
probability:
\begin{eqnarray}
  \label{eq:mfv}
  &&\Lambda > 5.5\, \mathrm{TeV} \quad \mathrm{(small}\, \tan \beta
  \mathrm{)}\,, \\
  &&\Lambda > 5.1\, \mathrm{TeV} \quad \mathrm{(large}\, \tan \beta
  \mathrm{)}\,.
\end{eqnarray}
The bound for large $\tan \beta$ comes from contributions proportional
to the same operator present in the SM.

As mentioned above, at very large $\tan \beta$ additional
contributions to $C_4(\Lambda)$ can be generated by Higgs exchange.
From these contributions, we obtain the following lower bound on the
scale $\Lambda$, which in this case is the mass of non-standard Higgs
bosons:
\begin{equation}
  \label{eq:chiggs}
  M_H > 5 \, \sqrt{(a_0+a_1)(a_0+a_2)}
  \left(
    \frac{\tan\beta}{50}
  \right)\,  \mathrm{TeV}\,.
\end{equation}
In any given model, one can specify the value of the $a_i$ couplings
and of $\tan \beta$ to obtain a lower bound on the non-standard Higgs
mass. The bound we obtained is in agreement with
Ref.~\cite{dambrosio}, taking into account the present experimental
information. If a non-standard Higgs boson is seen at hadron
colliders, this implies an upper bound on the $a_i$ couplings and/or
$\tan \beta$.

\begin{figure}[htb!]
\begin{center}
\includegraphics[width=0.45\textwidth]{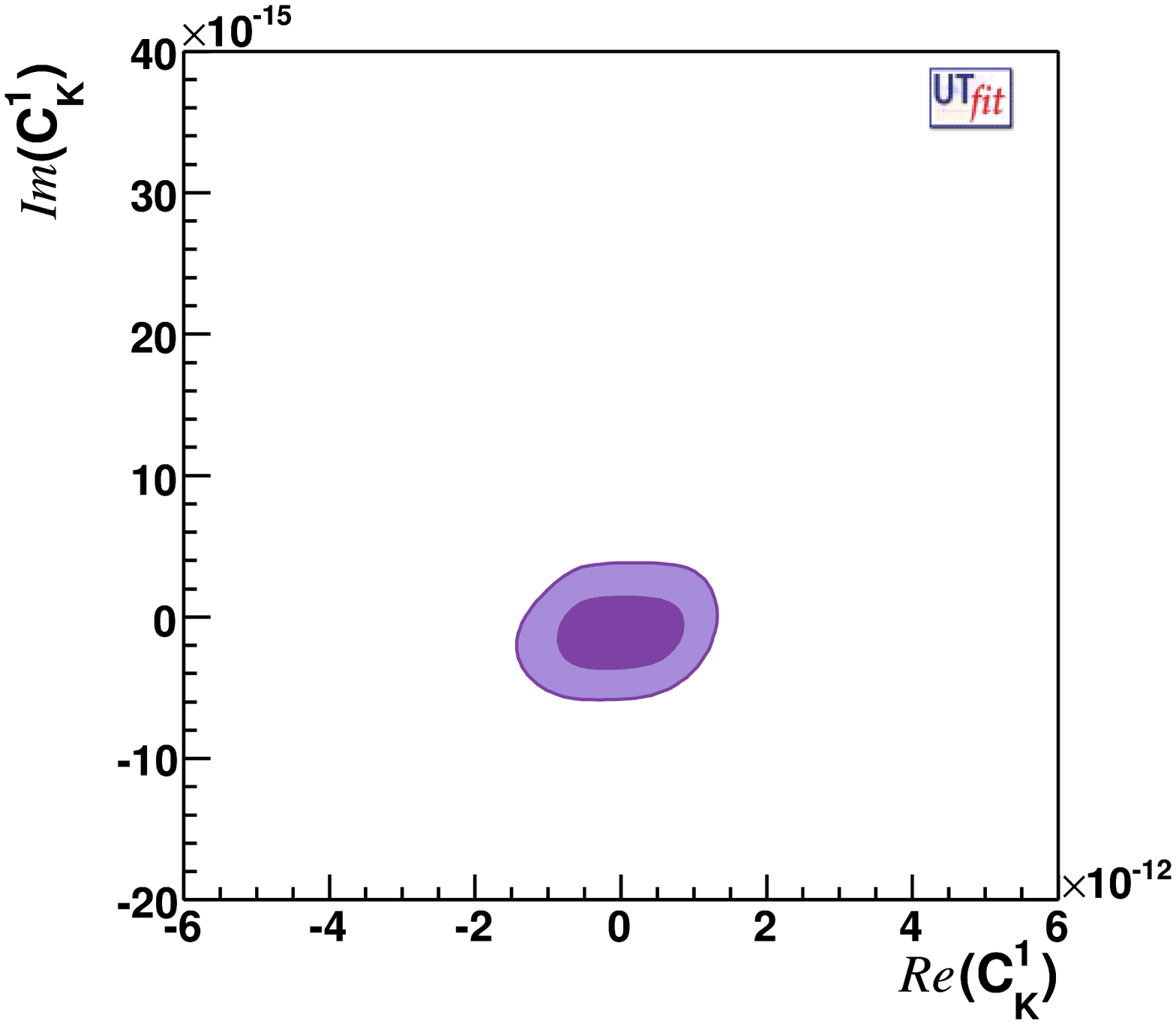}
\includegraphics[width=0.45\textwidth]{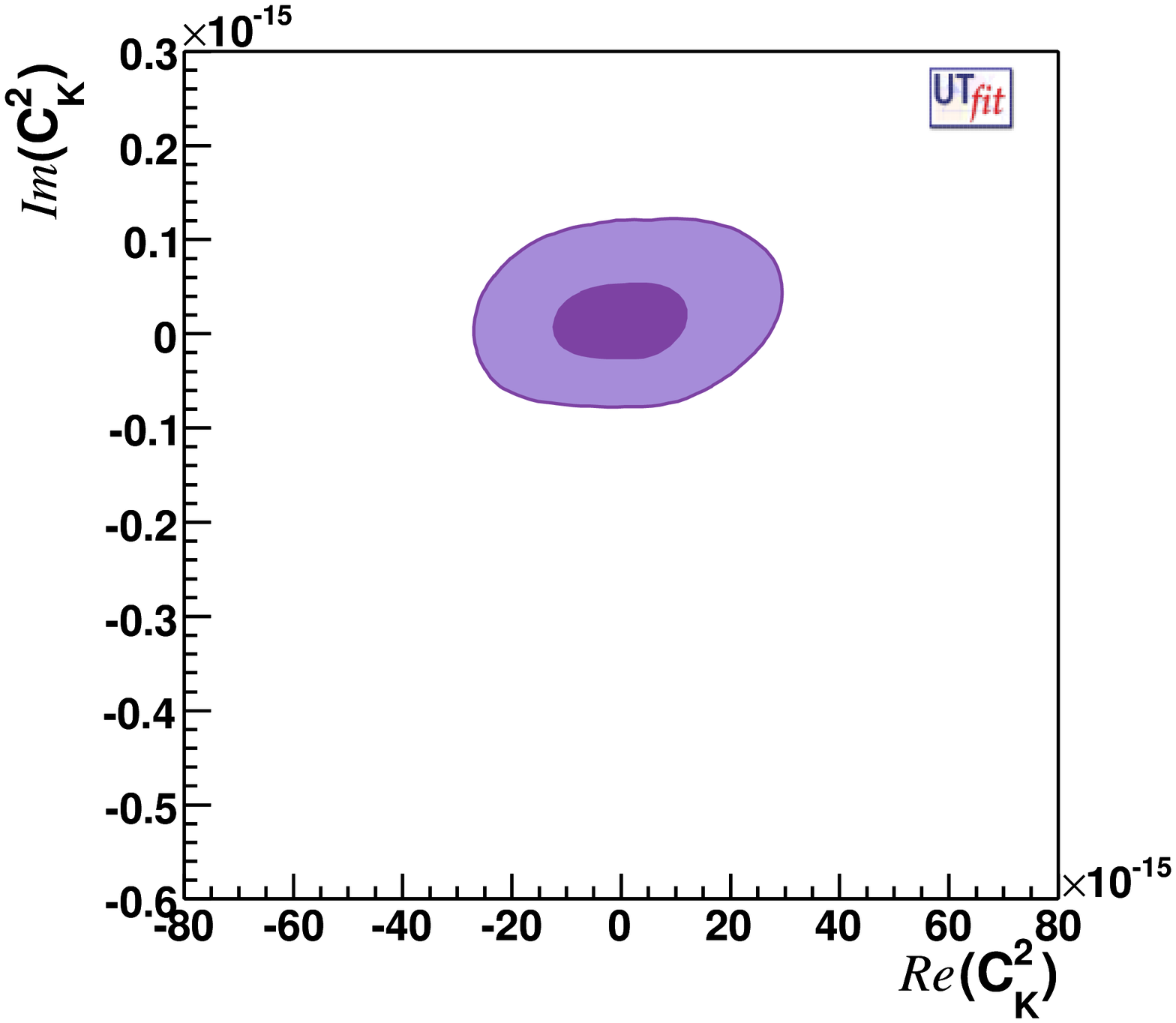}
\includegraphics[width=0.45\textwidth]{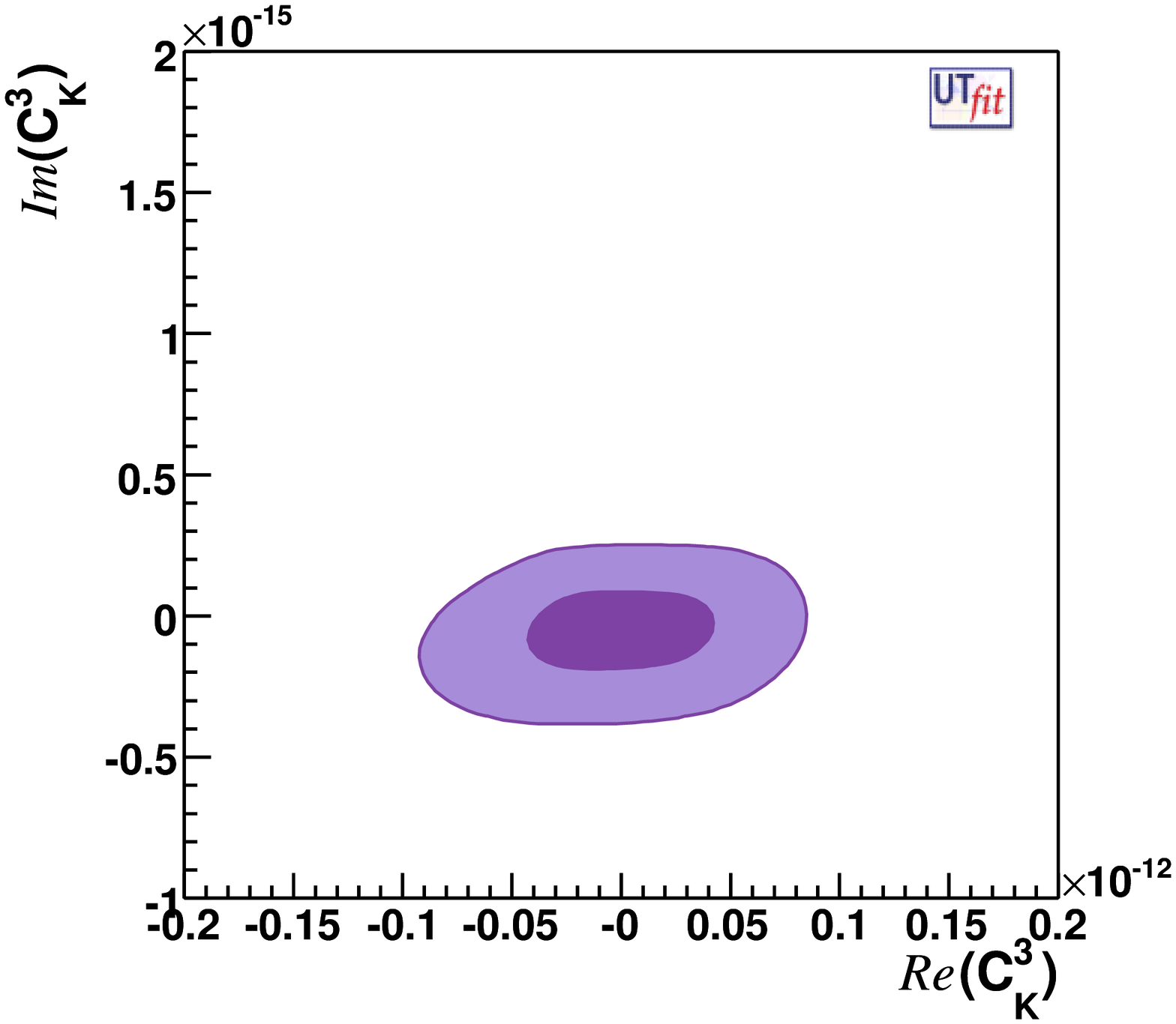}
\includegraphics[width=0.45\textwidth]{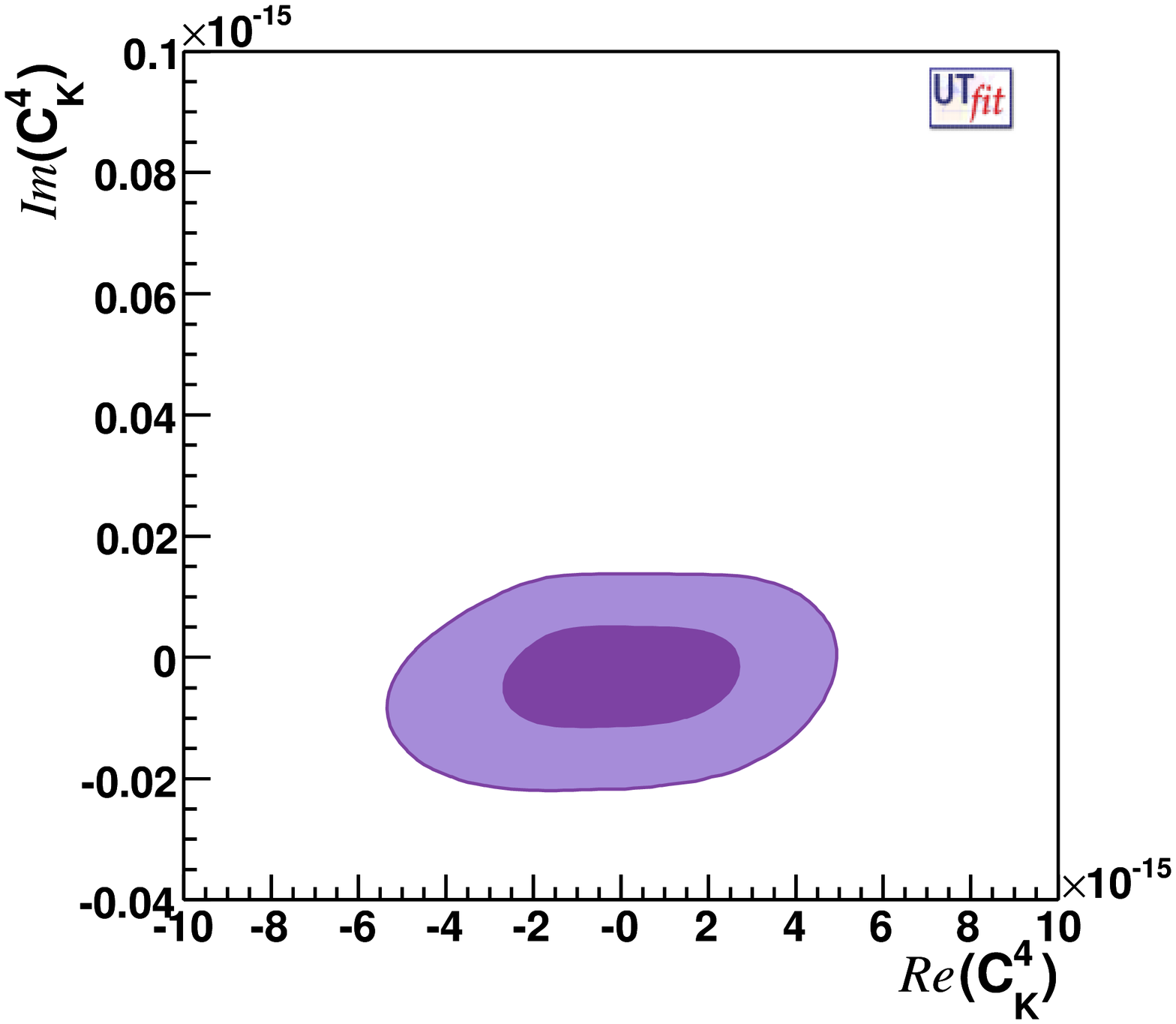}
\includegraphics[width=0.45\textwidth]{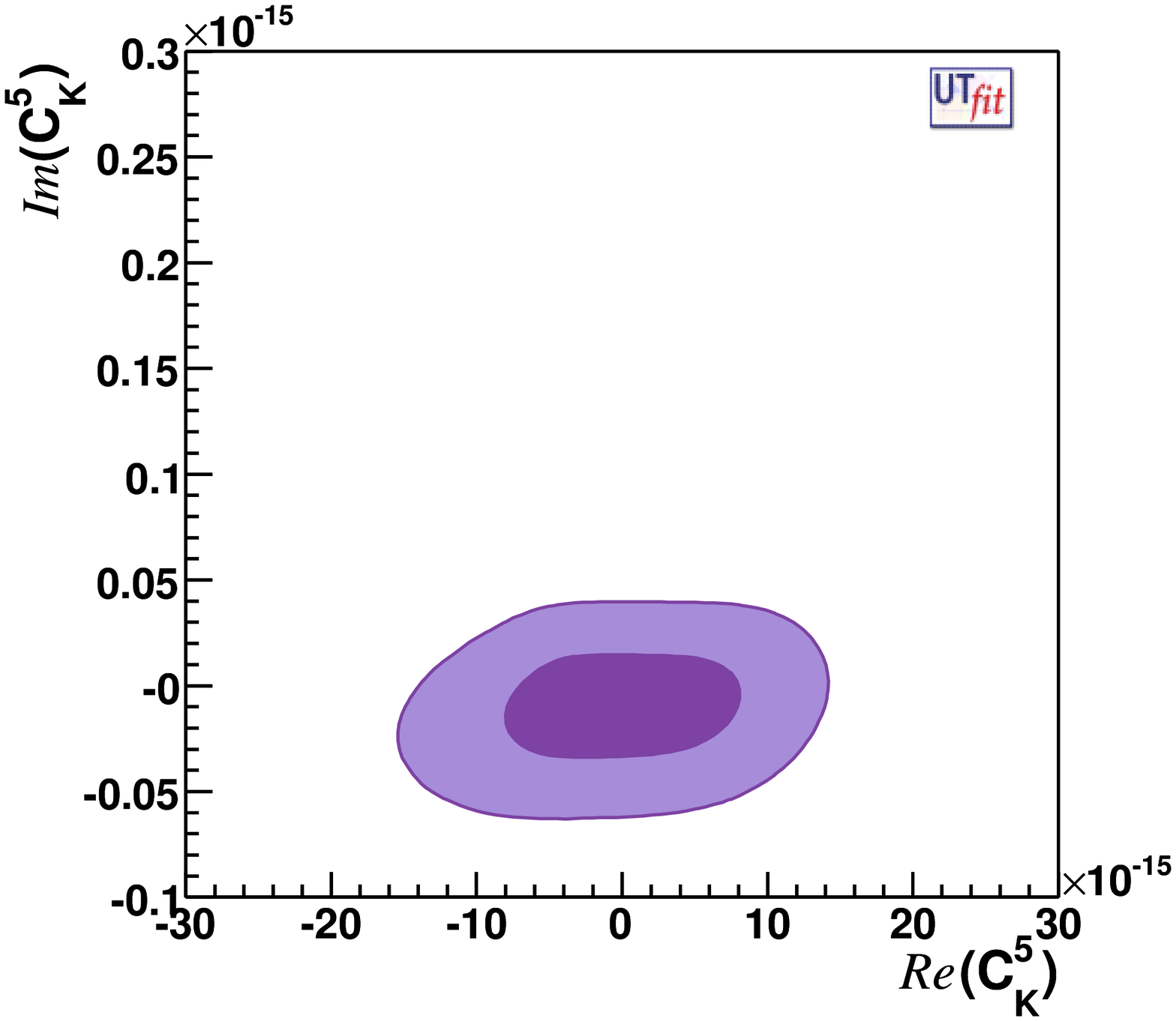}
\caption{Allowed ranges in the Re$C^i_K$-Im$C^i_K$ planes in
  GeV$^{-2}$. Light (dark) regions correspond to $95\%$ ($68\%$)
  probability regions.}
\label{fig:K}
\end{center}
\end{figure}

\begin{figure}[htb!]
\begin{center}
\includegraphics[width=0.45\textwidth]{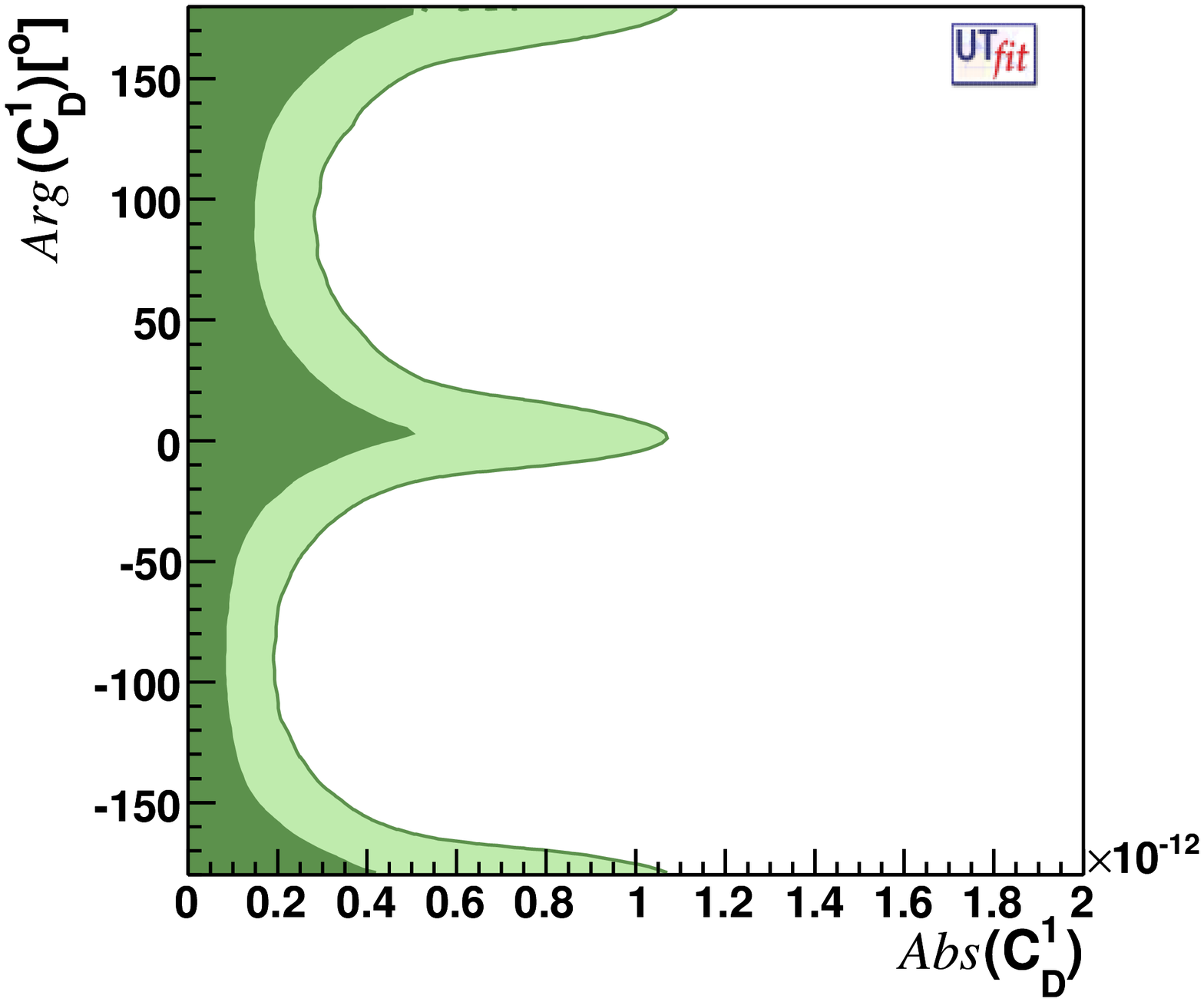}
\includegraphics[width=0.45\textwidth]{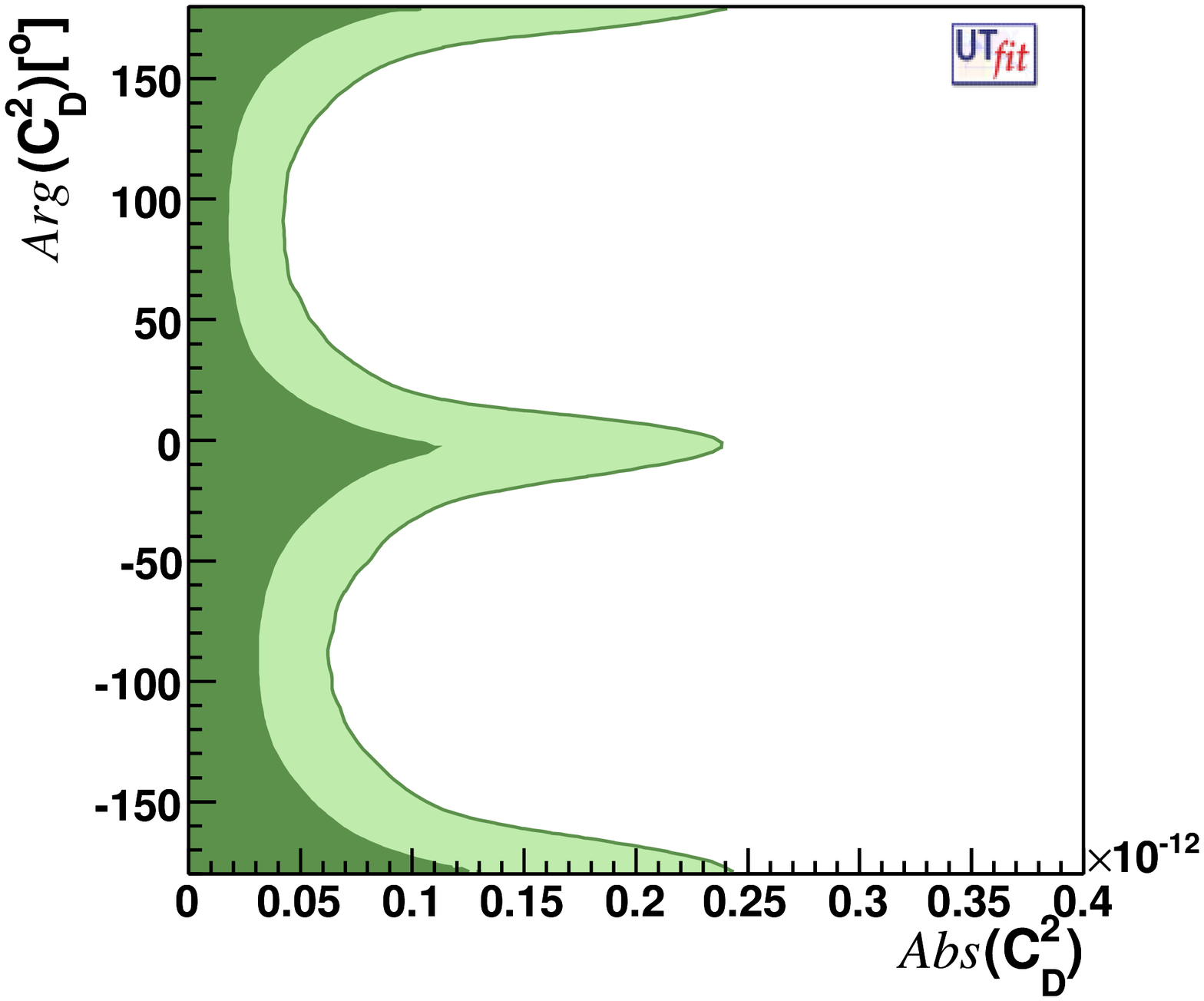}
\includegraphics[width=0.45\textwidth]{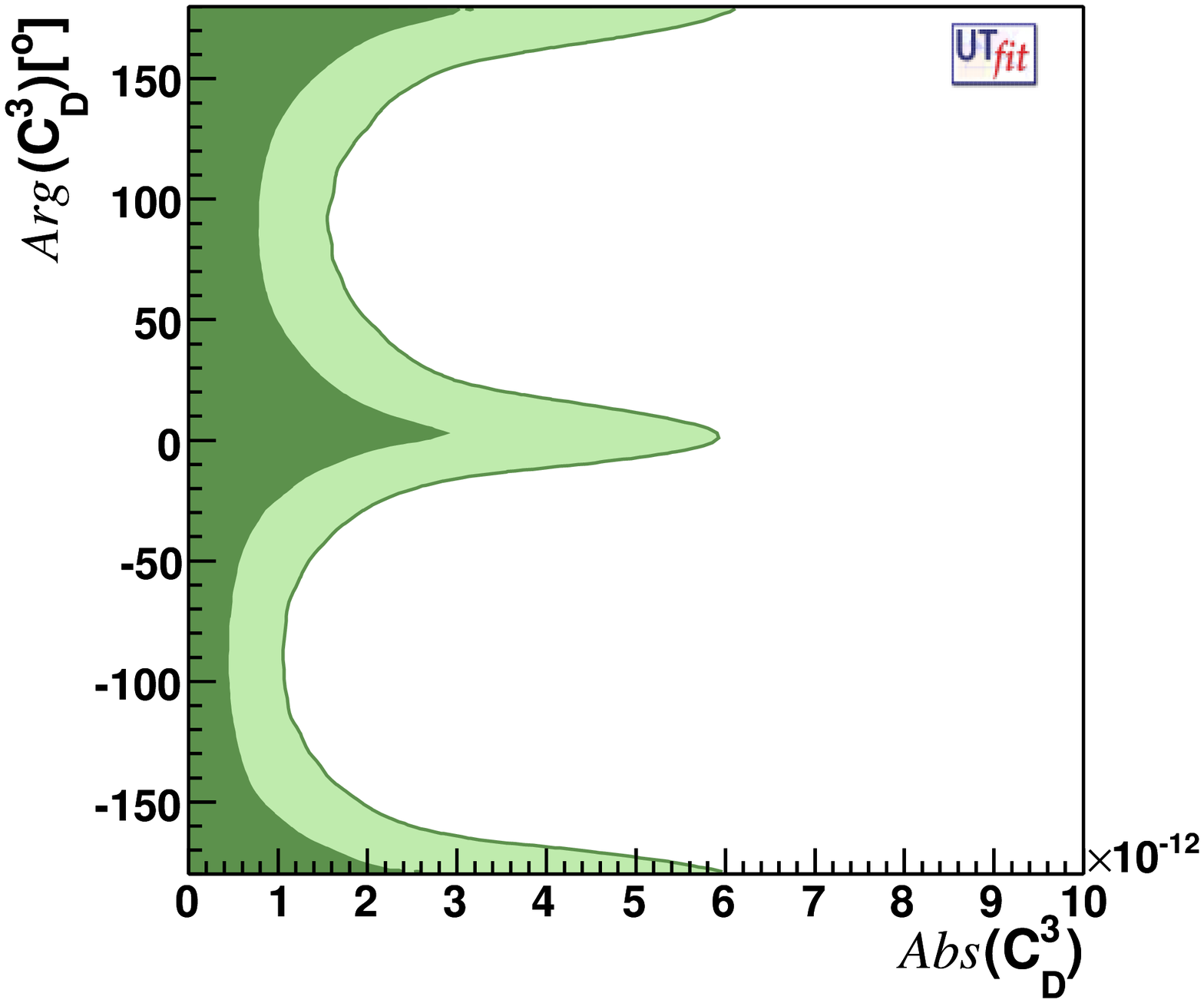}
\includegraphics[width=0.45\textwidth]{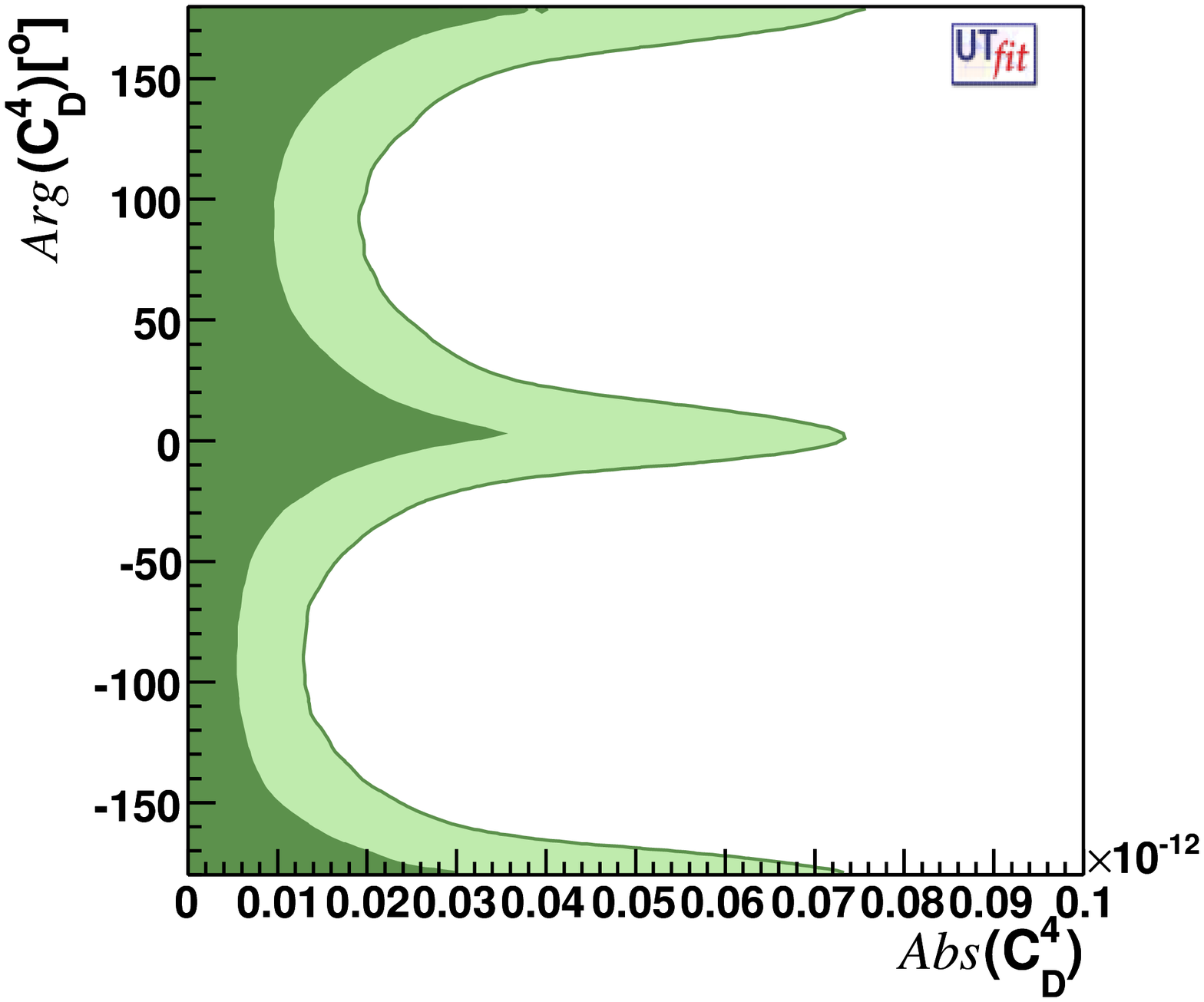}
\includegraphics[width=0.45\textwidth]{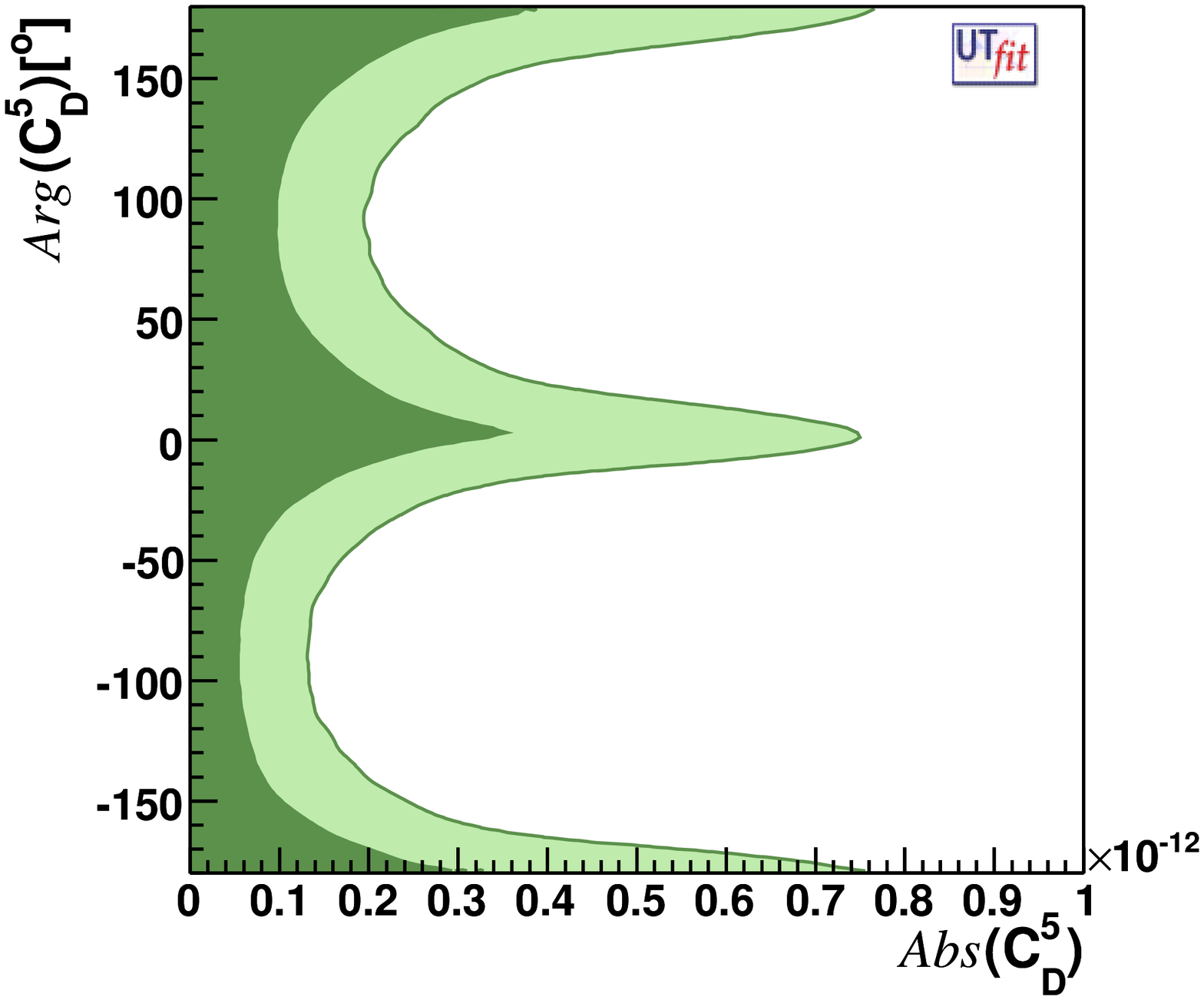}
\caption{Allowed ranges in the Abs$C^i_D$-Arg$C^i_D$ planes in
  GeV$^{-2}$. Light (dark) regions correspond to $95\%$ ($68\%$) 
  probability regions.}
\label{fig:D}
\end{center}
\end{figure}

\begin{figure}[htb!]
\begin{center}
\includegraphics[width=0.45\textwidth]{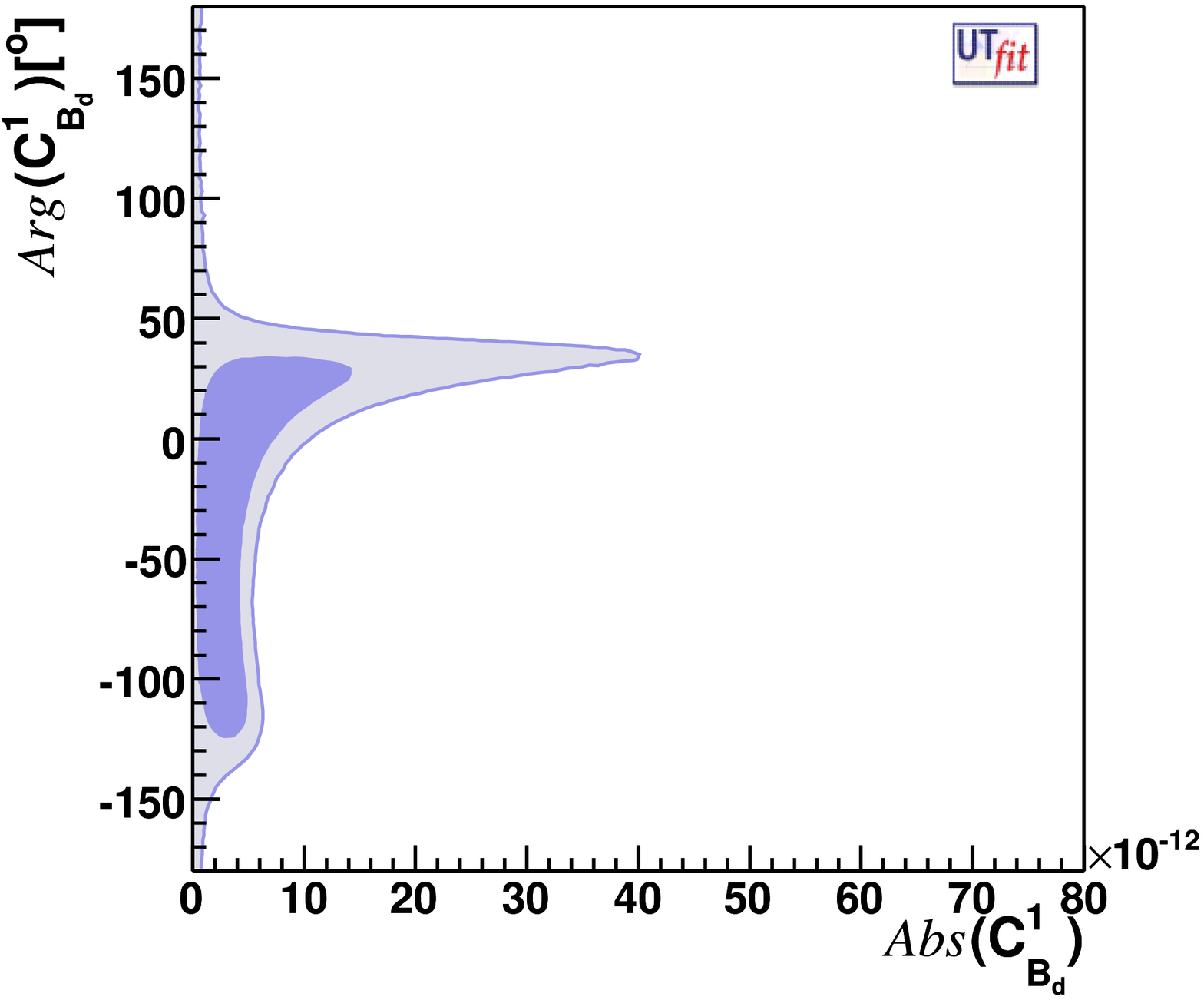}
\includegraphics[width=0.45\textwidth]{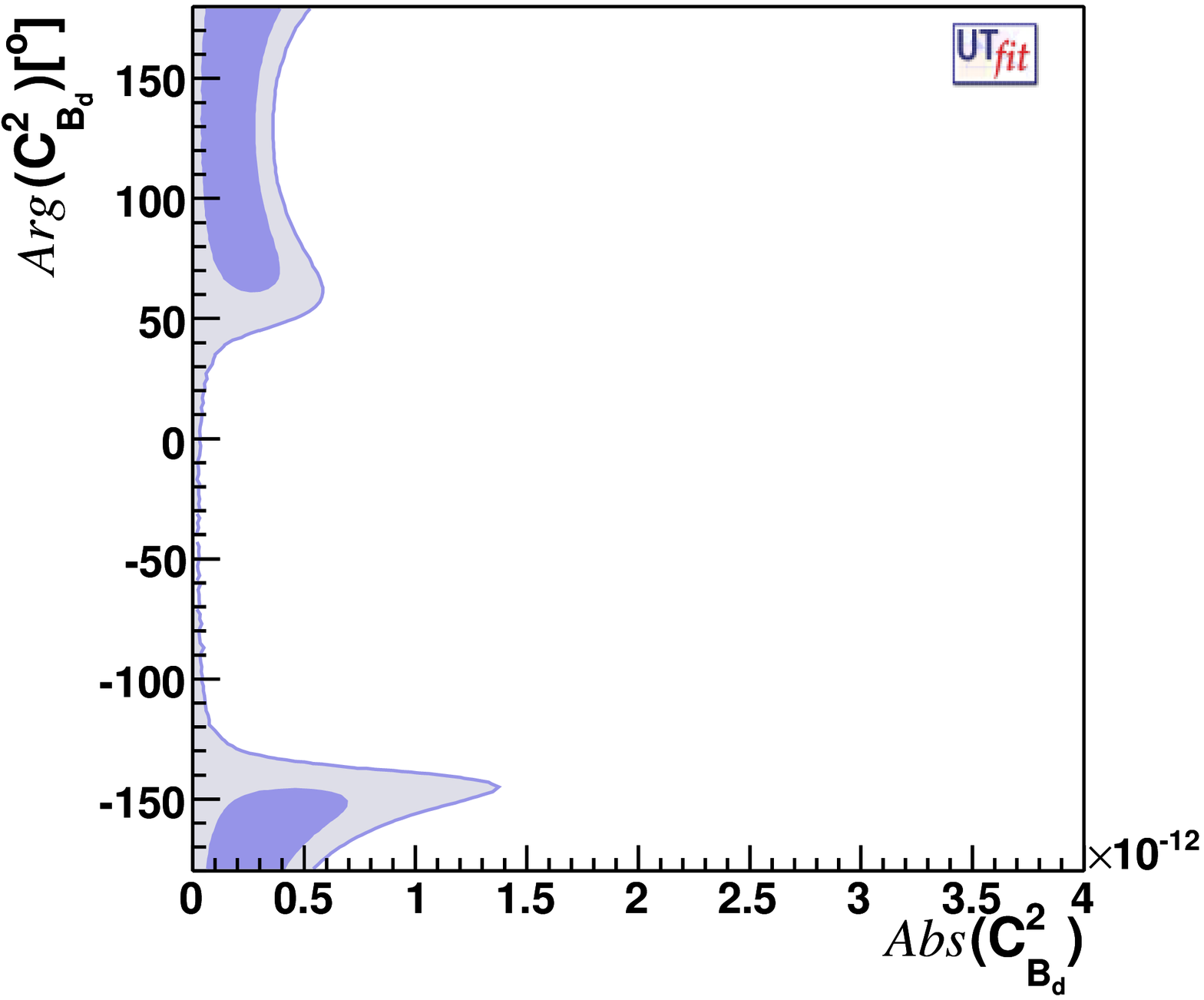}
\includegraphics[width=0.45\textwidth]{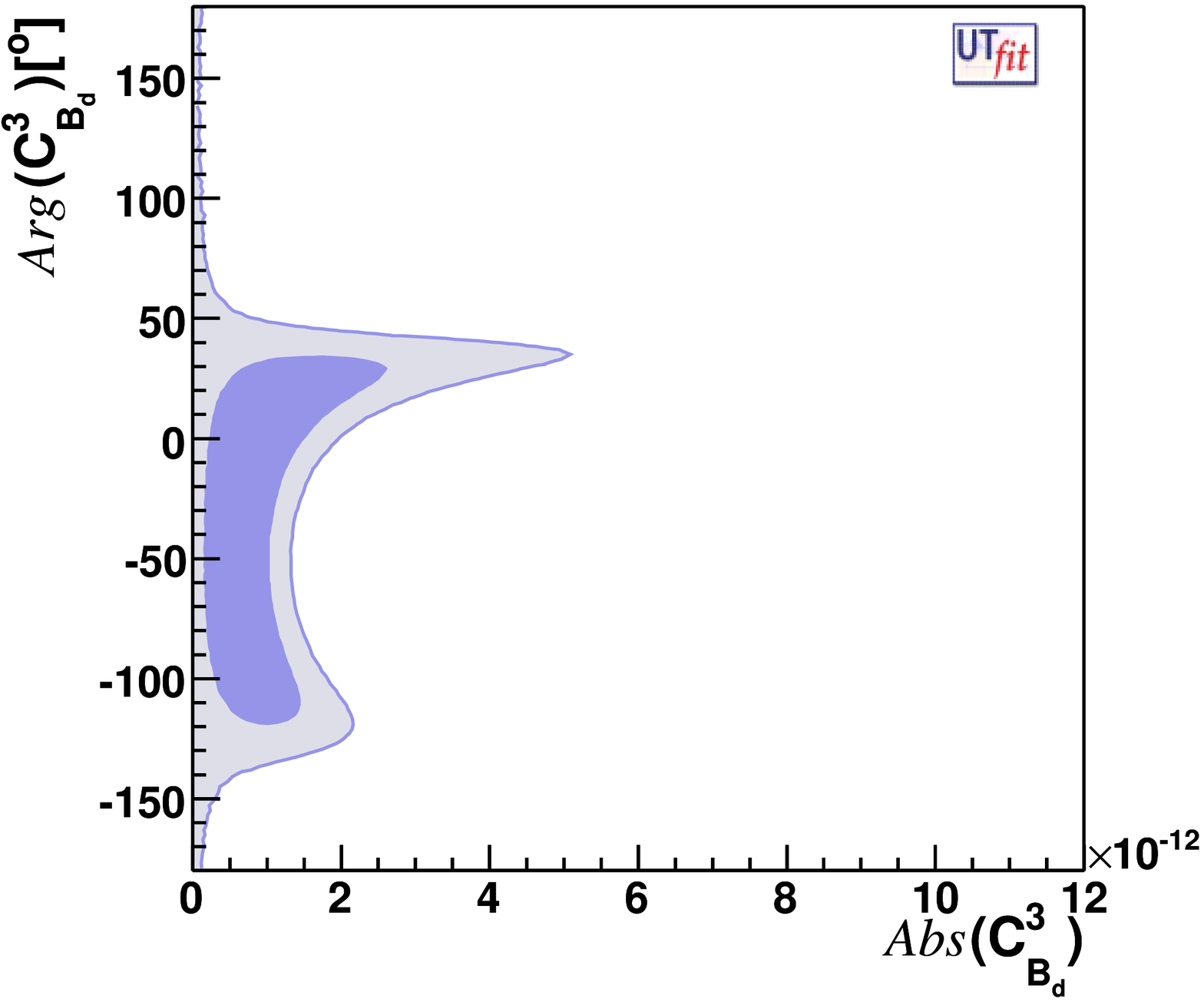}
\includegraphics[width=0.45\textwidth]{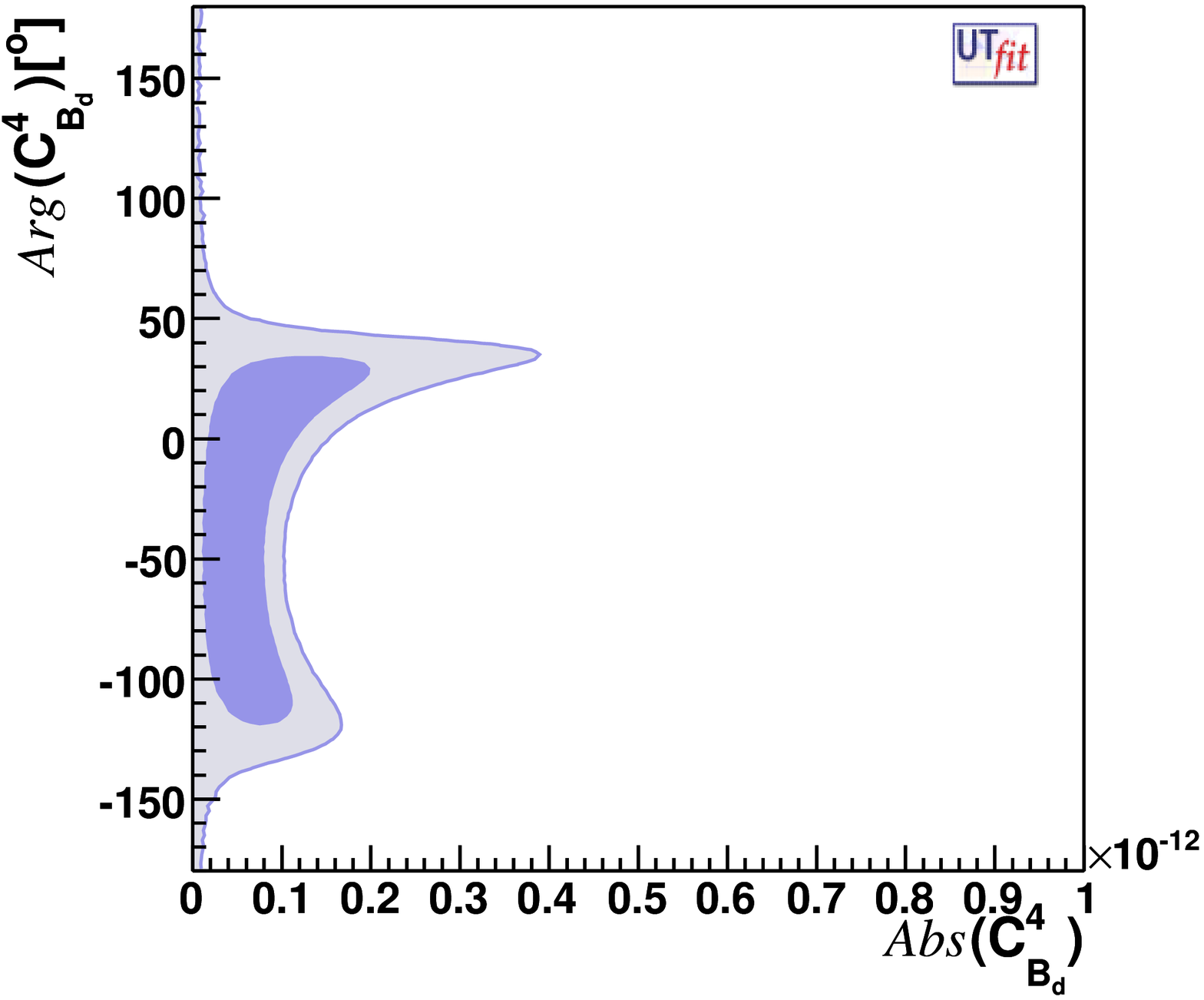}
\includegraphics[width=0.45\textwidth]{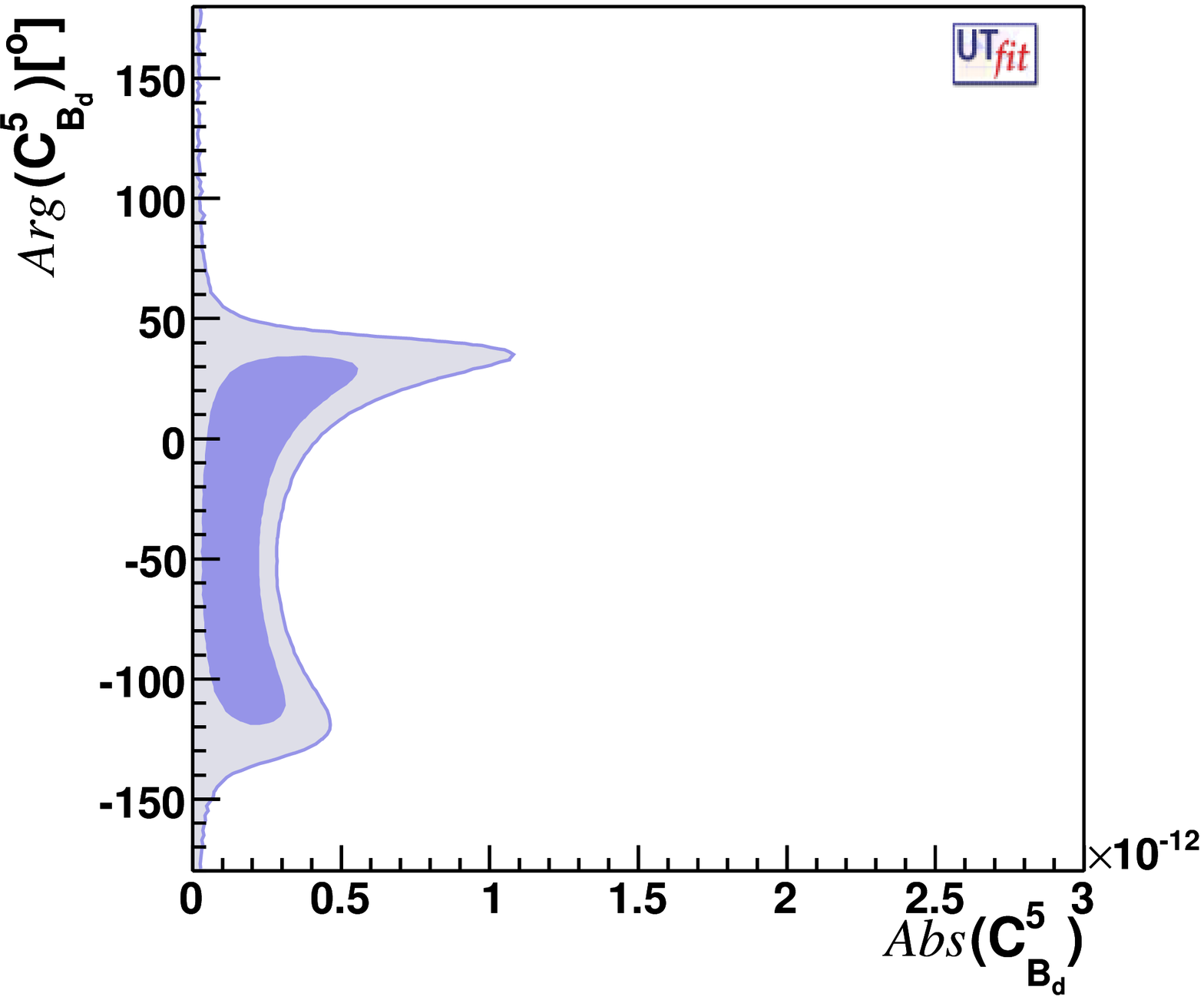}
\caption{Allowed ranges in the Abs$C^i_{B_d}$-Arg$C^i_{B_d}$ planes in
  GeV$^{-2}$. Light (dark) regions correspond to $95\%$
  ($68\%$) probability regions.}
\label{fig:Bd}
\end{center}
\end{figure}

\begin{figure}[htb!]
\begin{center}
\includegraphics[width=0.45\textwidth]{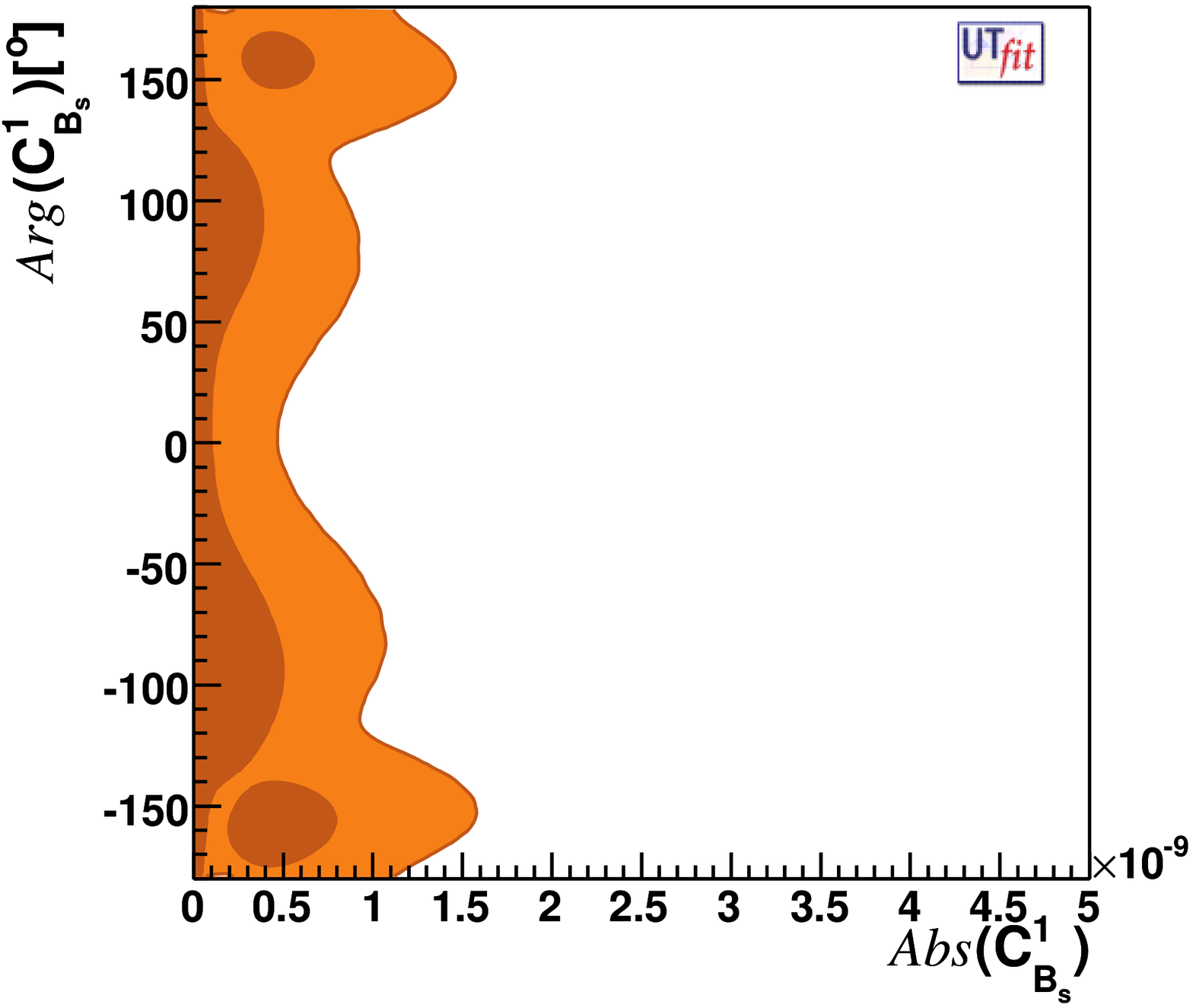}
\includegraphics[width=0.45\textwidth]{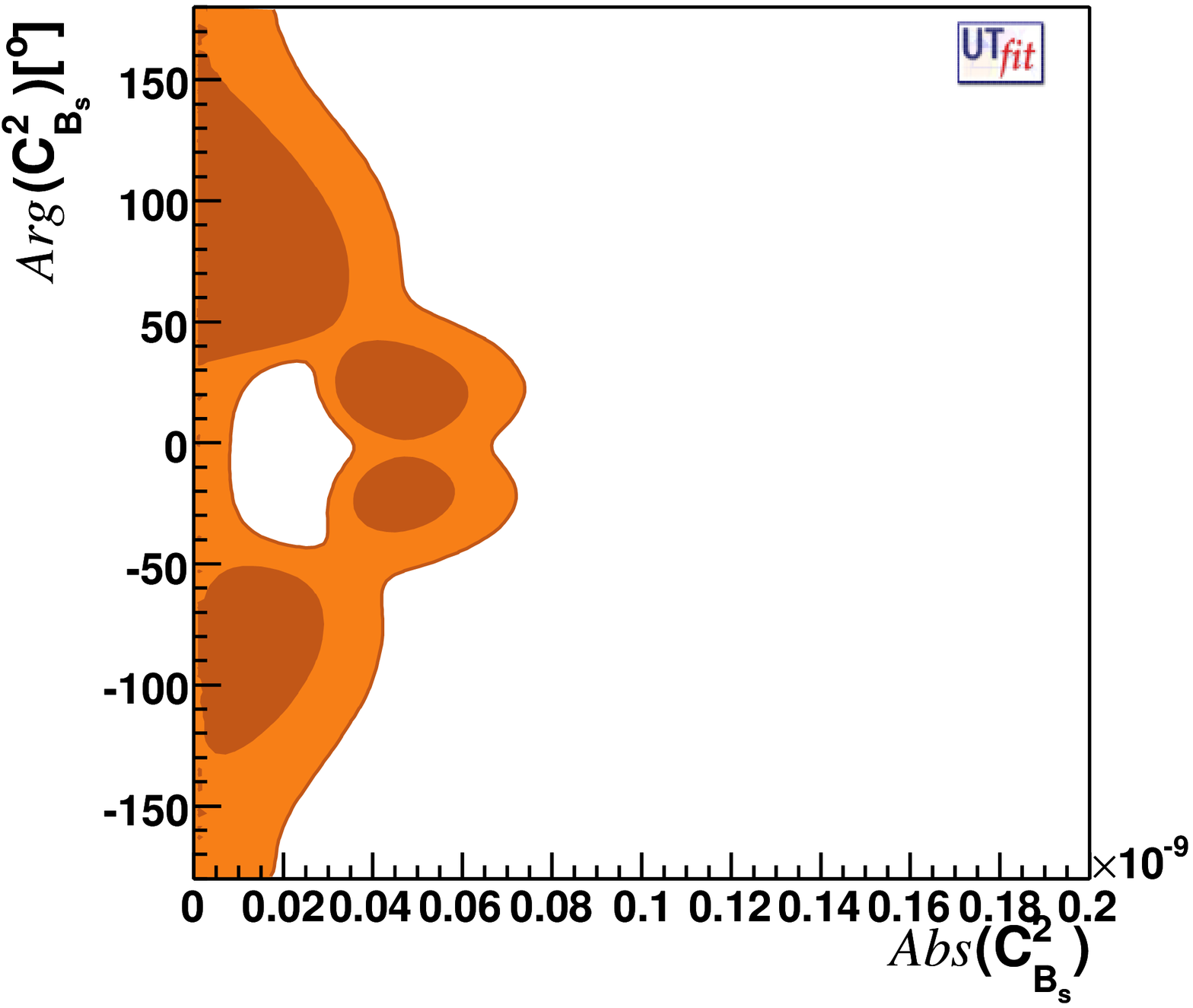}
\includegraphics[width=0.45\textwidth]{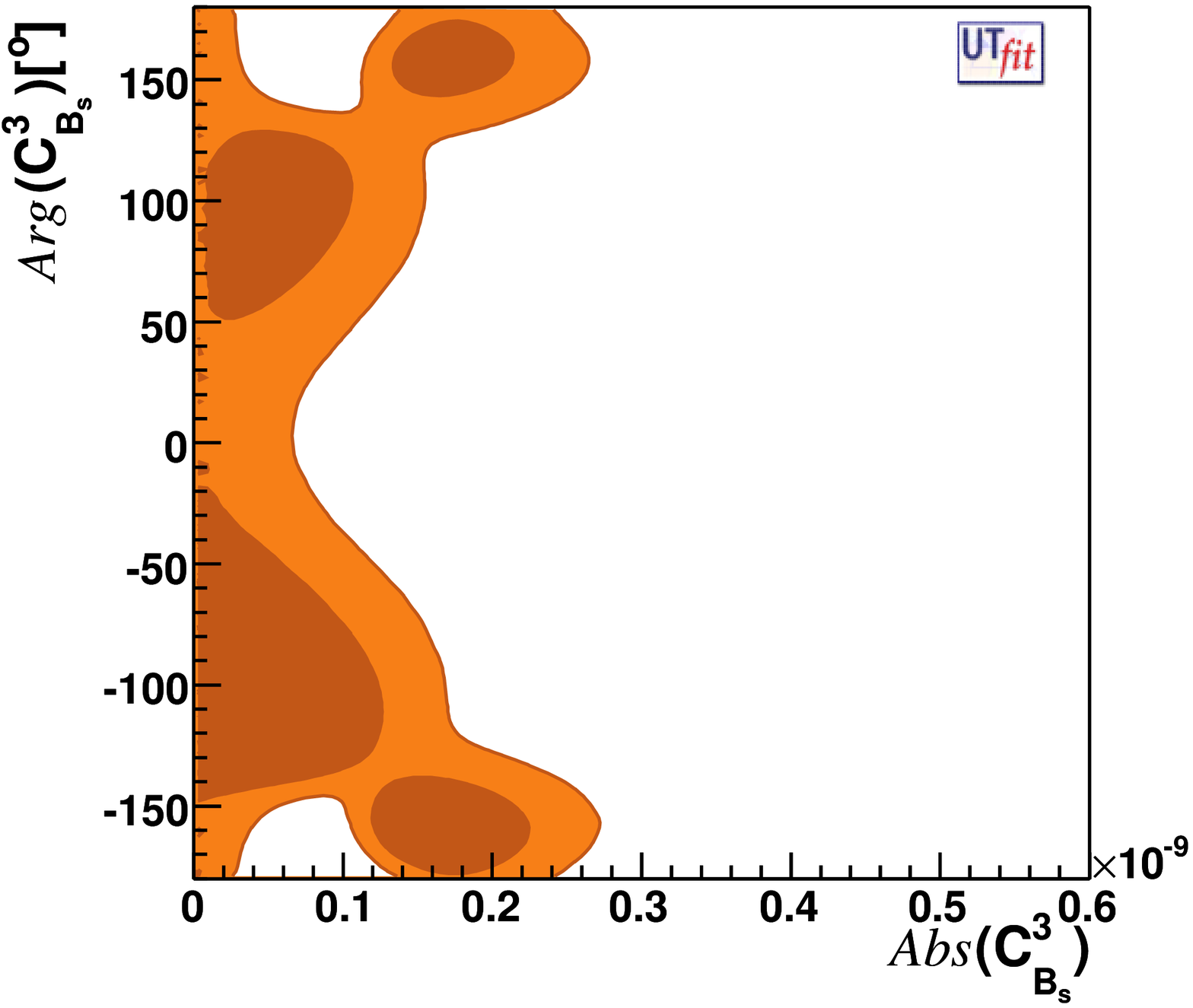}
\includegraphics[width=0.45\textwidth]{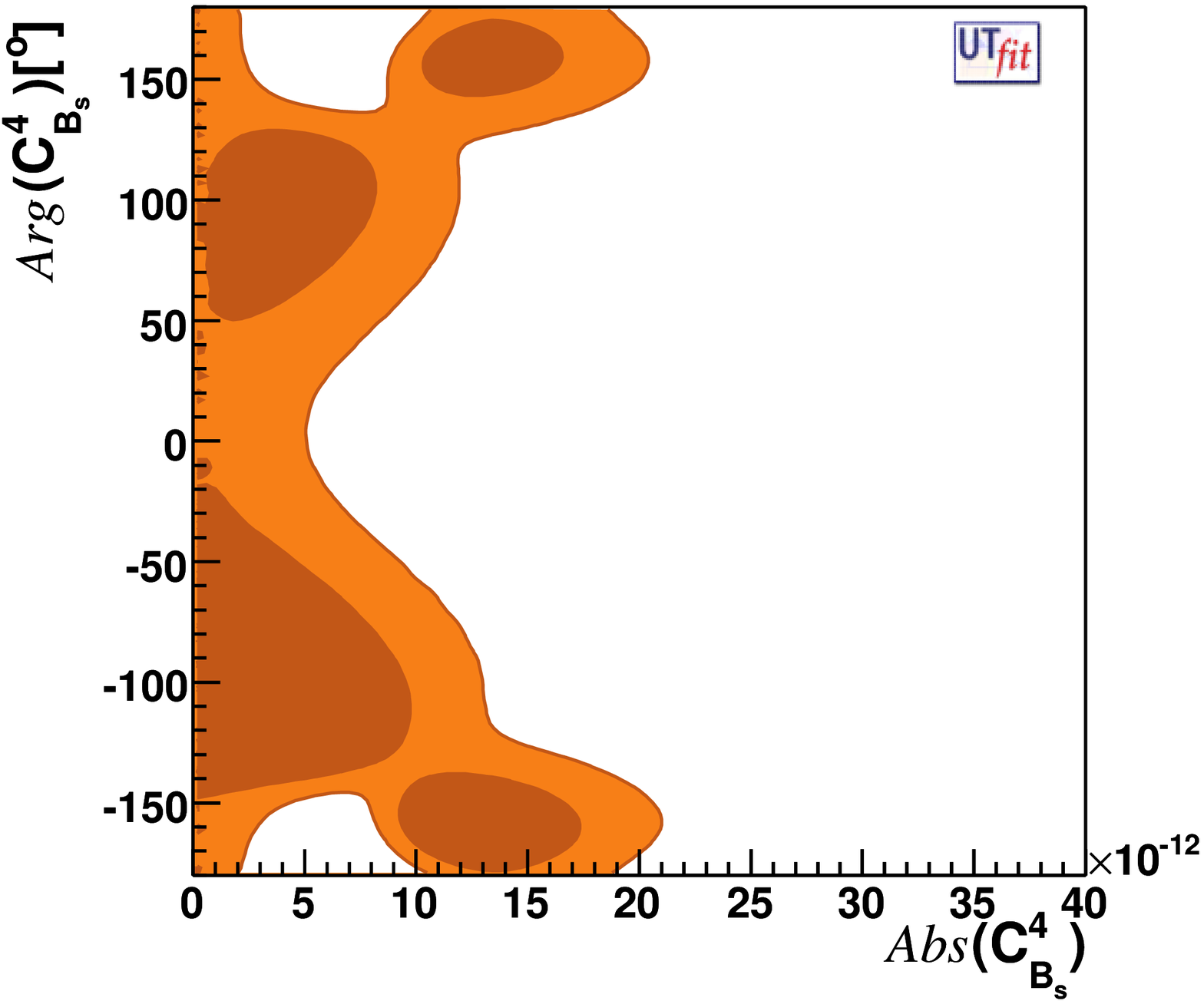}
\includegraphics[width=0.45\textwidth]{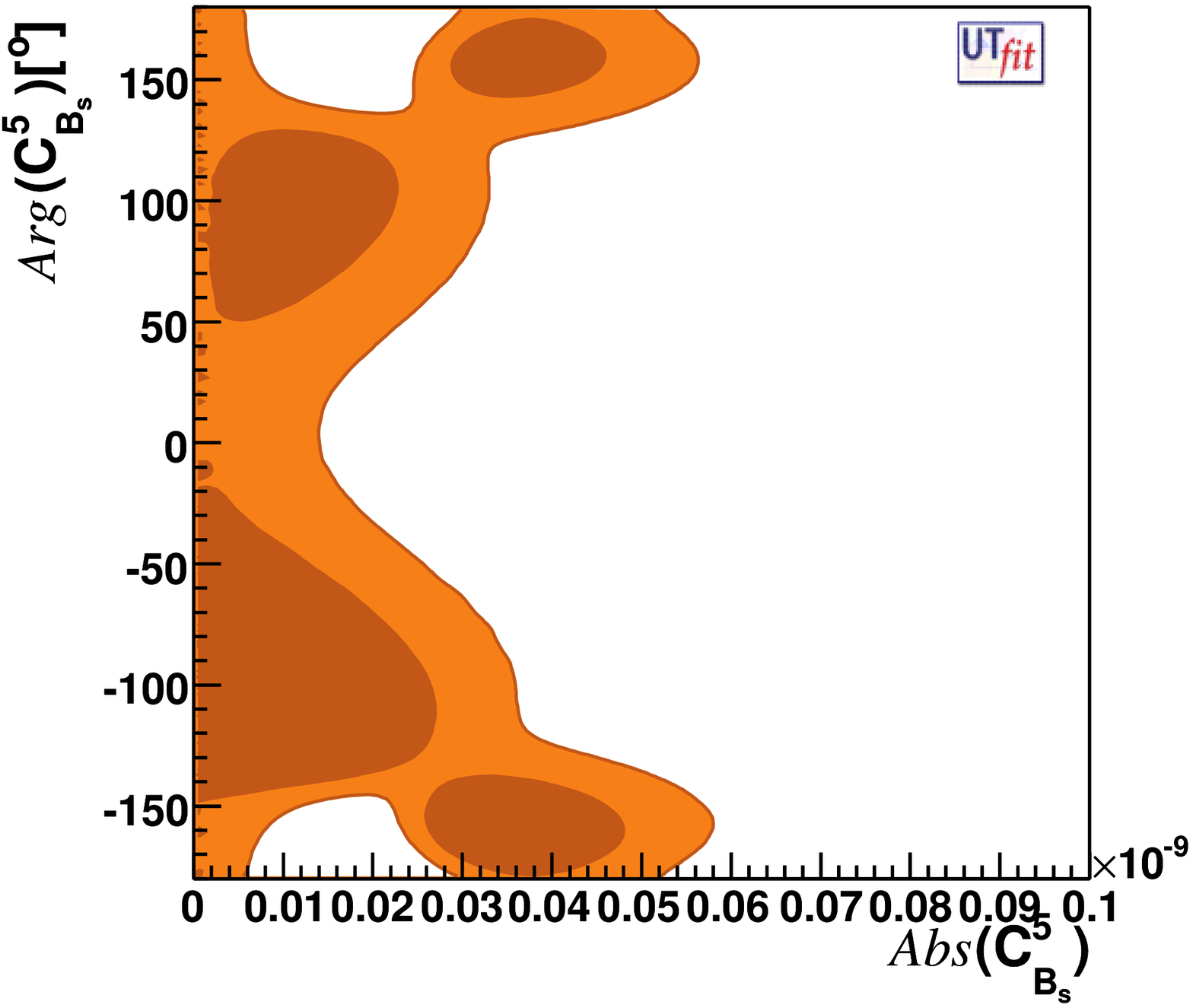}
\caption{Allowed ranges in the Abs$C^i_{B_s}$-Arg$C^i_{B_s}$ planes in
  GeV$^{-2}$. Light (dark) regions correspond to $95\%$
  ($68\%$) probability regions.}
\label{fig:Bs}
\end{center}
\end{figure}

\begin{table}[!h]
\begin{center}
\begin{tabular}{@{}cccc}
\hline\hline
Parameter &~~ $95\%$ allowed range~~ &~~ Lower limit on $\Lambda$ (TeV)&~~ Lower limit on $\Lambda$ (TeV)\\
          & (GeV$^{-2}$) &for arbitrary NP & for NMFV \\
\hline
Re$C_K^1$ & $[-9.6,9.6] \cdot 10^{-13}$ & $1.0 \cdot 10^{3}$& $0.35$ \\
Re$C_K^2$ & $[-1.8,1.9] \cdot 10^{-14}$ & $7.3 \cdot 10^{3}$& $2.0$ \\
Re$C_K^3$ & $[-6.0,5.6] \cdot 10^{-14}$ & $4.1 \cdot 10^{3}$& $1.1$ \\
Re$C_K^4$ & $[-3.6,3.6] \cdot 10^{-15}$ & $17 \cdot 10^{3}$ & $4.0$\\
Re$C_K^5$ & $[-1.0,1.0] \cdot 10^{-14}$ & $10 \cdot 10^{3}$ & $2.4$\\
\hline
Im$C_K^1$ & $[-4.4,2.8] \cdot 10^{-15}$ & $1.5 \cdot 10^{4}$& $5.6$ \\
Im$C_K^2$ & $[-5.1,9.3] \cdot 10^{-17}$ & $10 \cdot 10^{4}$ & $28$ \\
Im$C_K^3$ & $[-3.1,1.7] \cdot 10^{-16}$ & $5.7 \cdot 10^{4}$& $19$  \\
Im$C_K^4$ & $[-1.8,0.9] \cdot 10^{-17}$ & $24 \cdot 10^{4}$ & $62$ \\
Im$C_K^5$ & $[-5.2,2.8] \cdot 10^{-17}$ & $14 \cdot 10^{4}$ & $37$ \\
\hline
\hline
$|C_{D}^1|$ & $<7.2 \cdot 10^{-13}$ & $1.2 \cdot 10^{3}$& $0.40$\\
$|C_{D}^2|$ & $<1.6 \cdot 10^{-13}$ & $2.5 \cdot 10^{3}$& $0.82$\\
$|C_{D}^3|$ & $<3.9 \cdot 10^{-12}$ & $0.51\cdot 10^{3}$& $0.17$\\
$|C_{D}^4|$ & $<4.8 \cdot 10^{-14}$ & $4.6 \cdot 10^{3}$& $1.5$ \\
$|C_{D}^5|$ & $<4.8 \cdot 10^{-13}$ & $1.4 \cdot 10^{3}$& $0.47$\\
\hline
\hline
$|C_{B_d}^1|$ & $<2.3 \cdot 10^{-11}$ & $0.21 \cdot 10^{3}$& $1.8$\\
$|C_{B_d}^2|$ & $<7.2 \cdot 10^{-13}$ & $1.2 \cdot 10^{3}$ & $8.5$\\
$|C_{B_d}^3|$ & $<2.8 \cdot 10^{-12}$ & $0.60 \cdot 10^{3}$& $4.4$\\
$|C_{B_d}^4|$ & $<2.1 \cdot 10^{-13}$ & $2.2 \cdot 10^{3}$ & $14$ \\
$|C_{B_d}^5|$ & $<6.0 \cdot 10^{-13}$ & $1.3 \cdot 10^{3}$ & $8.8$\\
\hline
\hline
$|C_{B_s}^1|$ & $<1.1 \cdot 10^{-9}$ & $30$ & $1.3$\\
$|C_{B_s}^2|$ & $<5.6 \cdot 10^{-11}$ & $130$& $4.6$\\
$|C_{B_s}^3|$ & $<2.1 \cdot 10^{-10}$ & $70$ & $2.4$\\
$|C_{B_s}^4|$ & $<1.6 \cdot 10^{-11}$ & $250$& $7.9$\\
$|C_{B_s}^5|$ & $<4.5 \cdot 10^{-11}$ & $150$& $4.9$\\
\hline
\hline
\end{tabular}
\end{center}
\caption {$95\%$ probability range for
  $C(\Lambda)$ and the corresponding lower bounds on the NP
  scale $\Lambda$ for arbitrary NP flavour structure and for NMFV. See
  the text for 
  details.}
\label{tab:all}
\end{table}

In Fig.~\ref{fig:K} we present the allowed regions in the
Re$C^i$-Im$C^i$ planes for the $K^0$ sector, while in
Figs.~\ref{fig:D}-\ref{fig:Bs} we show the allowed regions in the
Abs$C^i$-Arg$C^i$ planes for the $D^0$, $B_d$ and $B_s$ sectors. All
coefficients are given in GeV$^{-2}$.  From these allowed regions we
obtain the $95\%$ probability regions for $C^i$ reported in the second
column of Tab.~\ref{tab:all}.  This result is completely
model-independent.  

Assuming strongly interacting and/or tree-level NP contributions with
generic flavour structure (\emph{i.e.} $L_i=\vert F_i\vert=1$), we can
translate the upper bounds on $C_i$ into the lower bounds on the NP
scale $\Lambda$ reported in the third column of Tab.~\ref{tab:all}.
As anticipated above, we see that in the $K^0$ sector all bounds from
non-standard operators are one order of magnitude stronger than the
bound from the SM operator, due to the chiral enhancement. In
addition, operator $Q_4$ has the strongest Renormalization Group (RG)
enhancement. In the $D^0$, $B_d$ and $B_s$ sectors, the chiral
enhancement is absent, but the RG enhancement is still effective.  The
overall constraint on the NP scale $\Lambda$ comes from Im$C^4_K$ and
reads, for strongly interacting and/or tree-level NP, $\alpha_s$ loop
mediated or $\alpha_W$ loop mediated respectively:
\begin{equation}
  \label{eq:GenDF2scale}
  \Lambda^\mathrm{GEN}_\mathrm{tree} > 2.4 \cdot 10^5
  \,\mathrm{TeV},\quad 
  \Lambda^\mathrm{GEN}_\mathrm{\alpha_s} > 2.4 \cdot
  10^4 \,\mathrm{TeV},\quad 
  \Lambda^\mathrm{GEN}_\mathrm{\alpha_W} > 8 \cdot
  10^3 \,\mathrm{TeV}. 
\end{equation}

Assuming strongly interacting and/or tree-level NP contributions with
NMFV flavour structure (\emph{i.e.} $L_i=1$ and $\vert F_i\vert =\vert
F_\mathrm{SM}\vert$), we can translate the
upper bounds on $C_i$ into the lower bounds on the NP scale $\Lambda$
reported in the fourth column of Tab.~\ref{tab:all}. The flavour
structure of NMFV models implies that the bounds from the four
sectors are all comparable, the strongest one being obtained from
Im$C^4_{K}$ (barring, as always, accidental cancellations):
\begin{equation}
  \label{eq:NMFVbound}
  \Lambda^\mathrm{NMFV}_\mathrm{tree} > 62
  \,\mathrm{TeV},\quad 
  \Lambda^\mathrm{NMFV}_\mathrm{\alpha_s} > 6.2\,\mathrm{TeV},\quad 
  \Lambda^\mathrm{NMFV}_\mathrm{\alpha_W} > 2\,\mathrm{TeV}. 
\end{equation}

Let us now comment on the possibility of direct detection of NP at
LHC, given the bounds we obtained. Clearly, a loop suppression is
needed in all scenarios to obtain NP scales that can be reached at the
LHC. For NMFV models, an $\alpha_W$ loop suppression might not be
sufficient, since the resulting NP scale is $2$ TeV. Of course, if
there is an accidental suppression of the NP contribution to
$\epsilon_K$, the scale for weak loop contributions might be as low as
$0.5$ TeV. The general model is out of reach even for $\alpha_W$ (or
stronger) loop suppression. For MFV models at large values of $\tan
\beta$, stringent constraints on the mass of the non-standard Higgs
bosons can be obtained. These particles may or may not be detectable
at the LHC depending on the actual value of $\tan \beta$. Finally, the
reader should keep in mind the possibility of accidental cancellations
among the contribution of different operators, which might weaken the
bounds we obtained.

\section{Conclusions}
\label{sec:concl}

\begin{figure}[htb]
  \centering
  \includegraphics[width=0.45\textwidth]{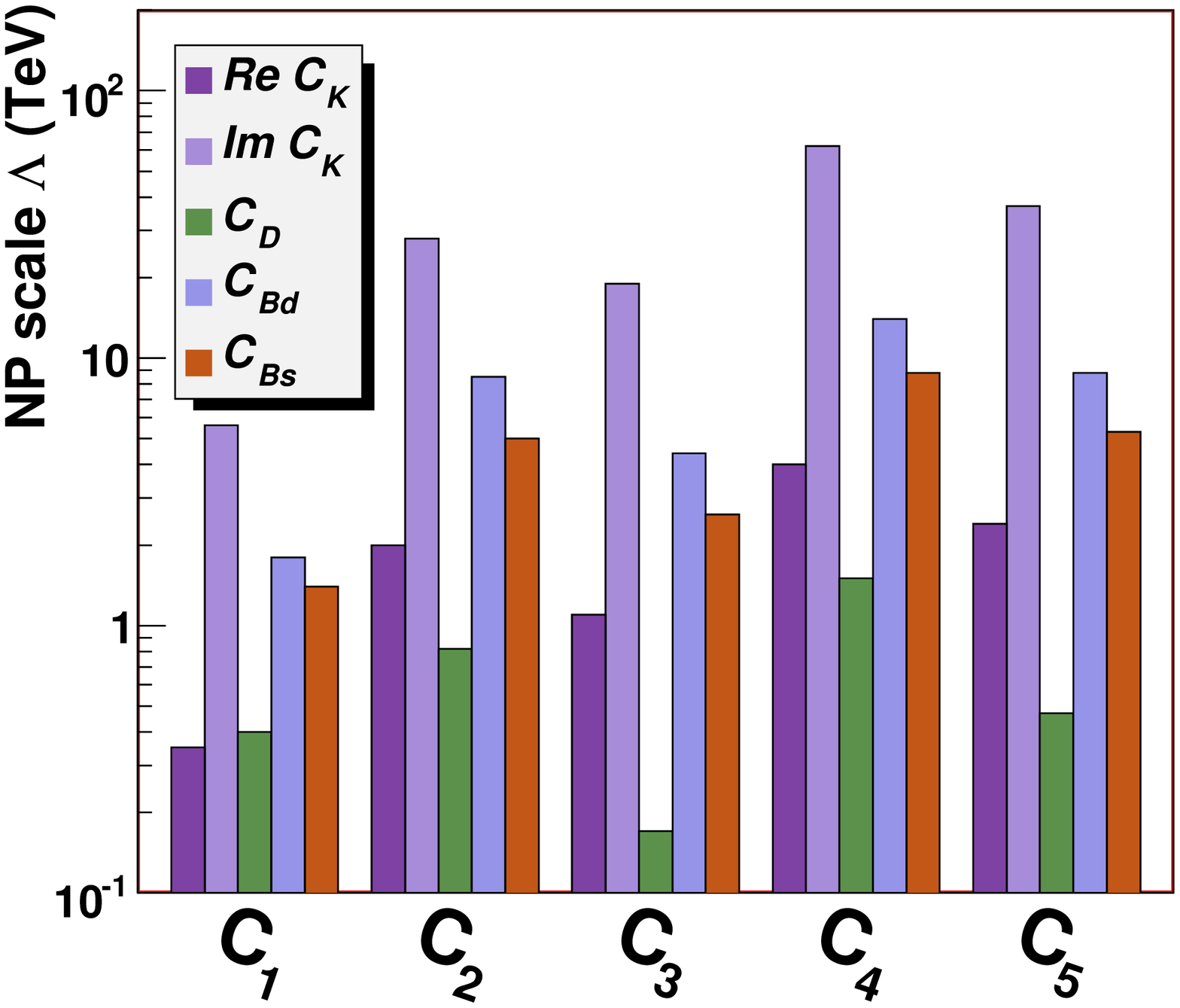}
  \includegraphics[width=0.45\textwidth]{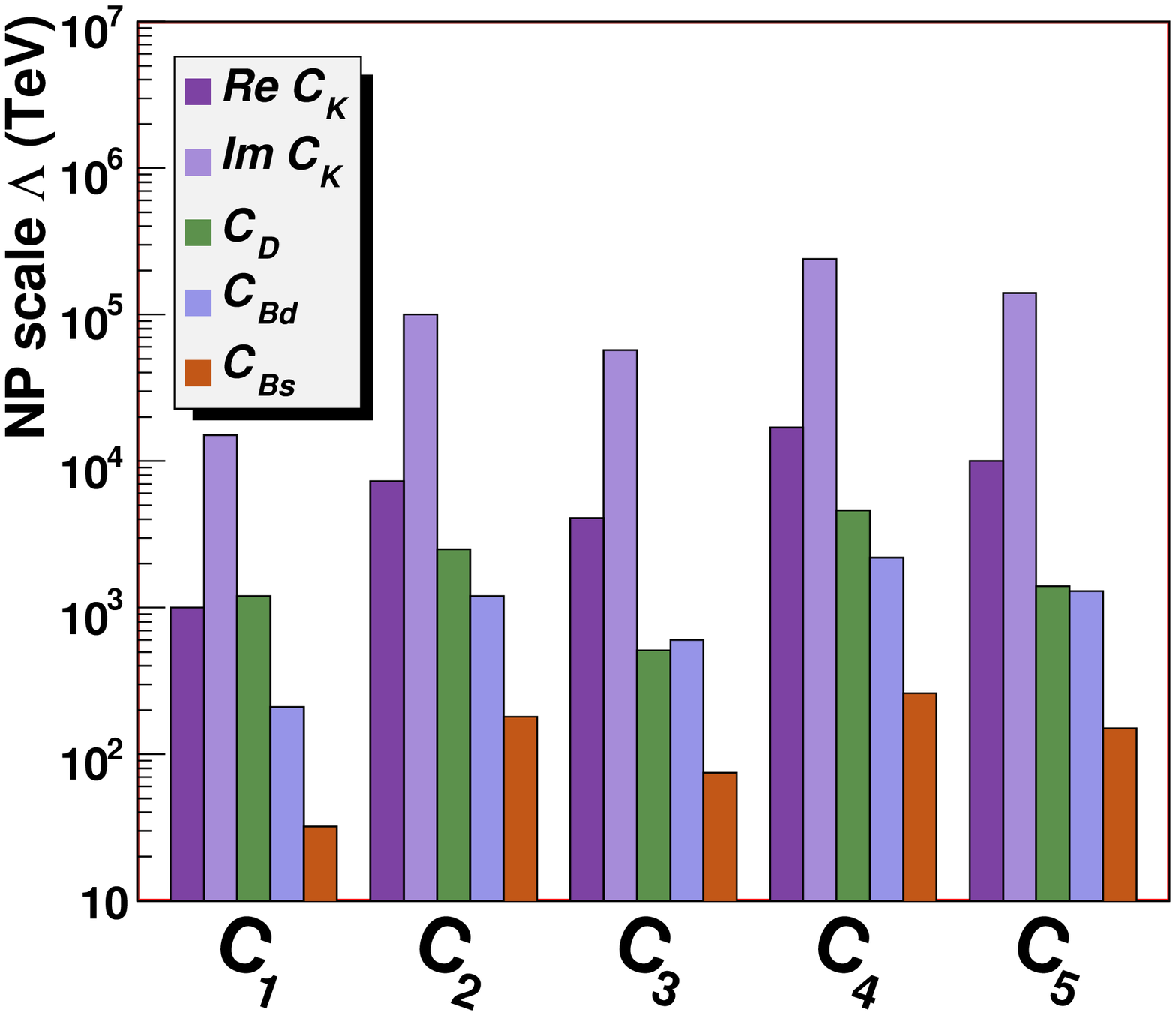}
  \caption{Summary of the $95 \%$ probability lower bound on the NP
    scale $\Lambda$ for strongly-interacting NP in NMFV
    (left) and general NP (right) scenarios.}
\label{fig:lambda}
\end{figure}

We have presented bounds on the NP scale $\Lambda$ obtained from an
operator analysis of $\Delta F=2$ processes, using the most recent
experimental measurements, the NLO formulae for the RG evolution and
the Lattice QCD results for the matrix elements. We have considered
four scenarios: MFV at small $\tan \beta$, MFV at large $\tan \beta$,
NMFV and general NP with arbitrary flavour structure.  The lower
bounds on the scale $\Lambda$ of strongly-interacting NP for NMFV and
general NP scenarios (barring accidental cancellations) are reported
in Fig.~\ref{fig:lambda}.  Taking the most stringent bound for each
scenario, we obtain the bounds given in Table~\ref{tab:summary}.

\begin{table}[h!]
\begin{center}
  \begin{tabular}{|c|c|c|c|}
    \hline
    Scenario &~ strong/tree ~&~~ $\alpha_s$ loop ~~& ~~$\alpha_W$ loop~~ \\
    MFV (small $\tan \beta$) & 5.5  & 0.5  & 0.2  \\
    MFV (large $\tan \beta$) & 5.1  & 0.5  & 0.2  \\
    ~$M_H$ in MFV at large $\tan \beta$~ & 
    \multicolumn{3}{c|}{$5\, \sqrt{(a_0+a_1)(a_0+a_2)}
      \left(
        \frac{\tan\beta}{50}
      \right)\,$  } \\
    NMFV  & 62  & 6.2  & 2  \\
    General  & 24000   & 2400  & 800  \\
    \hline
  \end{tabular}\\
\end{center}
\caption{Summary of the $95 \%$ probability lower bound on the NP
  scale $\Lambda$ (in TeV) for several possible flavour structures and loop
  suppressions.} 
\label{tab:summary}
\end{table}

We conclude that any model with strongly interacting NP and/or
tree-level contributions is beyond the reach of direct searches at the
LHC. Flavour and CP violation remain the main tool to constrain (or
detect) such NP models. Weakly-interacting extensions of the SM can be
accessible at the LHC provided that they enjoy a MFV-like suppression
of $\Delta F=2$ processes, or at least a NMFV-like suppression with an
additional depletion of the NP contribution to $\epsilon_K$.

\section*{Acknowledgments}

We thank Guennadi Borissov, Brendan Casey, Stefano Giagu, Marco Rescigno,
Andrzej Zieminski, and Jure Zupan for useful discussions.  We aknowledge
partial support from RTN European contracts MRTN-CT-2004-503369 ``The
Quest for Unification'', MRTN-CT-2006-035482 ``FLAVIAnet'' and
MRTN-CT-2006-035505 ``Heptools''.


\begin{thebibliography}{99}

\bibitem{susyjurassic}
  J.~R.~Ellis and D.~V.~Nanopoulos,
  Phys.\ Lett.\  B {\bf 110} (1982) 44;
  J.~F.~Donoghue, H.~P.~Nilles and D.~Wyler,
  Phys.\ Lett.\  B {\bf 128} (1983) 55;
  J.~M.~Frere and M.~Belen Gavela,
  Phys.\ Lett.\  B {\bf 132} (1983) 107;
  M.~J.~Duncan and J.~Trampetic,
  Phys.\ Lett.\  B {\bf 134} (1984) 439;
  J.~M.~Gerard, W.~Grimus, A.~Raychaudhuri and G.~Zoupanos,
  Phys.\ Lett.\  B {\bf 140} (1984) 349;
  J.~M.~Gerard, W.~Grimus and A.~Raychaudhuri,
  Phys.\ Lett.\  B {\bf 145} (1984) 400;
  J.~M.~Gerard, W.~Grimus, A.~Masiero, D.~V.~Nanopoulos and A.~Raychaudhuri,
  Nucl.\ Phys.\  B {\bf 253} (1985) 93.

\bibitem{LRjurassic}
  G.~Beall, M.~Bander and A.~Soni,
  Phys.\ Rev.\ Lett.\  {\bf 48} (1982) 848.

\bibitem{pellicanino}
  F.~Gabbiani and A.~Masiero,
  Nucl.\ Phys.\  B {\bf 322} (1989) 235.

\bibitem{pellicani}
  E.~Gabrielli, A.~Masiero and L.~Silvestrini,
  Phys.\ Lett.\  B {\bf 374} (1996) 80
  [arXiv:hep-ph/9509379];
  F.~Gabbiani, E.~Gabrielli, A.~Masiero and L.~Silvestrini,
  Nucl.\ Phys.\  B {\bf 477} (1996) 321
  [arXiv:hep-ph/9604387].

\bibitem{bagger}
  J.~A.~Bagger, K.~T.~Matchev and R.~J.~Zhang,
  Phys.\ Lett.\  B {\bf 412} (1997) 77
  [arXiv:hep-ph/9707225].

\bibitem{NLOgen} 
  M.~Ciuchini, E.~Franco, V.~Lubicz, G.~Martinelli,
  I.~Scimemi and L.~Silvestrini,
  Nucl.\ Phys.\  B {\bf 523} (1998) 501
  [arXiv:hep-ph/9711402];
  A.~J.~Buras, M.~Misiak and J.~Urban,
  Nucl.\ Phys.\  B {\bf 586} (2000) 397
  [arXiv:hep-ph/0005183].

\bibitem{BBSUSY}
  M.~Ciuchini, E.~Franco, D.~Guadagnoli, V.~Lubicz, V.~Porretti and L.~Silvestrini,
  JHEP {\bf 0609} (2006) 013
  [arXiv:hep-ph/0606197].

\bibitem{SUSYKKbar}
  M.~Ciuchini {\it et al.},
  JHEP {\bf 9810} (1998) 008
  [arXiv:hep-ph/9808328].

\bibitem{SUSYBBbar}
  D.~Becirevic {\it et al.},
  Nucl.\ Phys.\  B {\bf 634} (2002) 105
  [arXiv:hep-ph/0112303].

\bibitem{ckm}
  N.~Cabibbo,
  Phys.\ Rev.\ Lett.\  {\bf 10} (1963) 531;
  M.~Kobayashi and T.~Maskawa,
  Prog.\ Theor.\ Phys.\  {\bf 49} (1973) 652.

\bibitem{UTfitNP05}
  M.~Bona {\it et al.}  [UTfit Collaboration],
  JHEP {\bf 0603} (2006) 080
  [arXiv:hep-ph/0509219].

\bibitem{UTfitNP06}
  M.~Bona {\it et al.}  [UTfit Collaboration],
  Phys.\ Rev.\ Lett.\  {\bf 97} (2006) 151803
  [arXiv:hep-ph/0605213].

\bibitem{othernp}
  M.~Ciuchini, E.~Franco, F.~Parodi, V.~Lubicz, L.~Silvestrini and A.~Stocchi,
{\it In the Proceedings of 2nd Workshop on the CKM Unitarity Triangle, Durham, England, 5-9 Apr 2003, pp WG306}
  [arXiv:hep-ph/0307195];
  S.~Laplace, Z.~Ligeti, Y.~Nir and G.~Perez,
  Phys.\ Rev.\  D {\bf 65} (2002) 094040
  [arXiv:hep-ph/0202010];
  Z.~Ligeti,
  Int.\ J.\ Mod.\ Phys.\  A {\bf 20} (2005) 5105
  [arXiv:hep-ph/0408267];
  F.~J.~Botella, G.~C.~Branco, M.~Nebot and M.~N.~Rebelo,
  Nucl.\ Phys.\  B {\bf 725} (2005) 155
  [arXiv:hep-ph/0502133];
  L.~Silvestrini,
  Int.\ J.\ Mod.\ Phys.\  A {\bf 21} (2006) 1738
  [arXiv:hep-ph/0510077];
  F.~J.~Botella, G.~C.~Branco and M.~Nebot,
  Nucl.\ Phys.\  B {\bf 768} (2007) 1
  [arXiv:hep-ph/0608100].

\bibitem{noiddbar}
  M.~Ciuchini, E.~Franco, D.~Guadagnoli, V.~Lubicz, M.~Pierini, V.~Porretti and L.~Silvestrini,
  Phys.\ Lett.\  B {\bf 655} (2007) 162
  [arXiv:hep-ph/0703204].

\bibitem{altriddbar}
  Y.~Nir,
  JHEP {\bf 0705} (2007) 102
  [arXiv:hep-ph/0703235];
  P.~Ball,
  J.\ Phys.\ G {\bf 34} (2007) 2199
  [arXiv:0704.0786 [hep-ph]];
  E.~Golowich, J.~Hewett, S.~Pakvasa and A.~A.~Petrov,
  arXiv:0705.3650 [hep-ph].

\bibitem{mfv}
  E.~Gabrielli and G.~F.~Giudice,
  Nucl.\ Phys.\  B {\bf 433} (1995) 3
  [Erratum-ibid.\  B {\bf 507} (1997) 549]
  [arXiv:hep-lat/9407029];
  M.~Misiak, S.~Pokorski and J.~Rosiek,
  Adv.\ Ser.\ Direct.\ High Energy Phys.\  {\bf 15} (1998) 795
  [arXiv:hep-ph/9703442];
  M.~Ciuchini, G.~Degrassi, P.~Gambino and G.~F.~Giudice,
  Nucl.\ Phys.\  B {\bf 534} (1998) 3
  [arXiv:hep-ph/9806308];
  C.~Bobeth, M.~Bona, A.~J.~Buras, T.~Ewerth, M.~Pierini, L.~Silvestrini and A.~Weiler,
  Nucl.\ Phys.\  B {\bf 726} (2005) 252
  [arXiv:hep-ph/0505110];
  M.~Blanke, A.~J.~Buras, D.~Guadagnoli and C.~Tarantino,
  JHEP {\bf 0610} (2006) 003
  [arXiv:hep-ph/0604057].

\bibitem{uut}
  A.~J.~Buras, P.~Gambino, M.~Gorbahn, S.~Jager and L.~Silvestrini,
  Phys.\ Lett.\  B {\bf 500} (2001) 161
  [arXiv:hep-ph/0007085].

\bibitem{dambrosio}
  G.~D'Ambrosio, G.~F.~Giudice, G.~Isidori and A.~Strumia,
  Nucl.\ Phys.\  B {\bf 645} (2002) 155
  [arXiv:hep-ph/0207036].

\bibitem{papucci}
  K.~Agashe, M.~Papucci, G.~Perez and D.~Pirjol,
  arXiv:hep-ph/0509117.

\bibitem{barbierietal}
  B.~Grinstein and M.~B.~Wise,
  Phys.\ Lett.\  B {\bf 265} (1991) 326;
  A.~Belyaev and R.~Rosenfeld,
  Mod.\ Phys.\ Lett.\  A {\bf 14} (1999) 397
  [arXiv:hep-ph/9805253];
  R.~Barbieri and A.~Strumia,
  Phys.\ Lett.\  B {\bf 462} (1999) 144
  [arXiv:hep-ph/9905281];
  R.~Barbieri, A.~Pomarol, R.~Rattazzi and A.~Strumia,
  Nucl.\ Phys.\  B {\bf 703} (2004) 127
  [arXiv:hep-ph/0405040].

\bibitem{soni}
  K.~Agashe, G.~Perez and A.~Soni,
  Phys.\ Rev.\ Lett.\  {\bf 93} (2004) 201804
  [arXiv:hep-ph/0406101];
  K.~Agashe, G.~Perez and A.~Soni,
  Phys.\ Rev.\  D {\bf 71} (2005) 016002
  [arXiv:hep-ph/0408134].

\bibitem{dmsCDF}
  A.~Abulencia {\it et al.}  [CDF Collaboration],
  Phys.\ Rev.\ Lett.\  {\bf 97} (2006) 242003
  [arXiv:hep-ex/0609040].

\bibitem{ASLD0}
  V.~M.~Abazov {\it et al.}  [D0 Collaboration],
  Phys.\ Rev.\ Lett.\  {\bf 98} (2007) 151801
  [arXiv:hep-ex/0701007].

\bibitem{ACHD0}
  V.~M.~Abazov {\it et al.}  [D0 Collaboration],
  Phys.\ Rev.\  D {\bf 74} (2006) 092001
  [arXiv:hep-ex/0609014].

\bibitem{ASLCDF} CDF Collaboration, CDF note 9015,\\
  \texttt{http://www-cdf.fnal.gov/physics/new/bottom/070816.blessed-acp-bsemil/}. 

\bibitem{tauBsflavspec}
  D.~Buskulic {\it et al.}  [ALEPH Collaboration],
  Phys.\ Lett.\  B {\bf 377} (1996) 205;
  F.~Abe {\it et al.}  [CDF Collaboration],
  Phys.\ Rev.\  D {\bf 59} (1999) 032004
  [arXiv:hep-ex/9808003];
  P.~Abreu {\it et al.}  [DELPHI Collaboration],
  Eur.\ Phys.\ J.\  C {\bf 16} (2000) 555
  [arXiv:hep-ex/0107077];
  K.~Ackerstaff {\it et al.}  [OPAL Collaboration],
  Phys.\ Lett.\  B {\bf 426} (1998) 161
  [arXiv:hep-ex/9802002];
  V.~M.~Abazov {\it et al.}  [D0 Collaboration],
  Phys.\ Rev.\ Lett.\  {\bf 97} (2006) 241801
  [arXiv:hep-ex/0604046];
  CDF Collaboration, CDF note 7386,\\
  \texttt{http://www-cdf.fnal.gov/physics/new/bottom/050303.blessed-bhadlife/};\\
  CDF Collaboration, CDF note 7757,\\ \texttt{http://www-cdf.fnal.gov/physics/new/bottom/050707.blessed-bs-semi\_life/};
  E.~Barberio {\it et al.}  [Heavy Flavor Averaging Group (HFAG)],
  arXiv:hep-ex/0603003.

\bibitem{DGoGCDF}
  D.~E.~Acosta {\it et al.}  [CDF Collaboration],
  Phys.\ Rev.\ Lett.\  {\bf 94} (2005) 101803
  [arXiv:hep-ex/0412057].

\bibitem{DGoGD0}
  V.~M.~Abazov {\it et al.}  [D0 Collaboration],
  Phys.\ Rev.\ Lett.\  {\bf 98} (2007) 121801
  [arXiv:hep-ex/0701012].

\bibitem{noi}
  M.~Ciuchini {\it et al.},
  JHEP {\bf 0107} (2001) 013
  [arXiv:hep-ph/0012308].

\bibitem{dighe}
  A.~S.~Dighe, I.~Dunietz and R.~Fleischer,
  Eur.\ Phys.\ J.\  C {\bf 6} (1999) 647
  [arXiv:hep-ph/9804253].

\bibitem{digheold}
  A.~S.~Dighe, I.~Dunietz, H.~J.~Lipkin and J.~L.~Rosner,
  Phys.\ Lett.\  B {\bf 369} (1996) 144
  [arXiv:hep-ph/9511363].

\bibitem{nierste}
  A.~Lenz and U.~Nierste,
  JHEP {\bf 0706} (2007) 072
  [arXiv:hep-ph/0612167].

\bibitem{kramer}
  G.~Kramer and W.~F.~Palmer,
  Phys.\ Rev.\  D {\bf 45}, 193 (1992).

\bibitem{neubert}
  M.~Neubert,
  Phys.\ Rept.\  {\bf 245}, 259 (1994)
  [arXiv:hep-ph/9306320].

\bibitem{beneke}
  M.~Beneke, J.~Rohrer and D.~Yang,
  Nucl.\ Phys.\  B {\bf 774}, 64 (2007)
  [arXiv:hep-ph/0612290].

\bibitem{babarjpsikst}
  B.~Aubert {\it et al.}  [BABAR Collaboration],
  Phys.\ Rev.\  D {\bf 71} (2005) 032005
  [arXiv:hep-ex/0411016].

\bibitem{tauBsCP}
  F.~Abe {\it et al.}  [CDF Collaboration],
  Phys.\ Rev.\  D {\bf 57} (1998) 5382.

\bibitem{teoTauBs}
  K.~Hartkorn and H.~G.~Moser,
  Eur.\ Phys.\ J.\  C {\bf 8} (1999) 381.

\bibitem{nir}
  Y.~Grossman, Y.~Nir and G.~Raz,
  Phys.\ Rev.\ Lett.\  {\bf 97} (2006) 151801
  [arXiv:hep-ph/0605028].

\bibitem{ddbarexp}
  E.~M.~Aitala {\it et al.}  [E791 Collaboration],
  Phys.\ Rev.\ Lett.\  {\bf 77} (1996) 2384
  [arXiv:hep-ex/9606016];
  J.~M.~Link {\it et al.}  [FOCUS Collaboration],
  Phys.\ Lett.\  B {\bf 485} (2000) 62
  [arXiv:hep-ex/0004034];
  K.~Abe {\it et al.}  [Belle Collaboration],
  Phys.\ Rev.\ Lett.\  {\bf 88} (2002) 162001
  [arXiv:hep-ex/0111026];
  S.~E.~Csorna {\it et al.}  [CLEO Collaboration],
  Phys.\ Rev.\  D {\bf 65} (2002) 092001
  [arXiv:hep-ex/0111024];
  B.~Aubert {\it et al.}  [BABAR Collaboration],
  Phys.\ Rev.\ Lett.\  {\bf 91} (2003) 121801
  [arXiv:hep-ex/0306003];
  B.~Aubert {\it et al.}  [BABAR Collaboration],
  Phys.\ Rev.\  D {\bf 70} (2004) 091102
  [arXiv:hep-ex/0408066];
  C.~Cawlfield {\it et al.}  [CLEO Collaboration],
  Phys.\ Rev.\  D {\bf 71} (2005) 077101
  [arXiv:hep-ex/0502012];
  D.~M.~Asner {\it et al.}  [CLEO Collaboration],
  Phys.\ Rev.\  D {\bf 72} (2005) 012001
  [arXiv:hep-ex/0503045];
  U.~Bitenc {\it et al.}  [Belle Collaboration],
  Phys.\ Rev.\  D {\bf 72} (2005) 071101
  [arXiv:hep-ex/0507020];
  D.~M.~Asner {\it et al.}  [CLEO Collaboration],
  Int.\ J.\ Mod.\ Phys.\  A {\bf 21} (2006) 5456
  [arXiv:hep-ex/0607078];
  B.~Aubert {\it et al.}  [BABAR Collaboration],
  arXiv:hep-ex/0607090;
  B.~Aubert {\it et al.}  [BABAR Collaboration],
  Phys.\ Rev.\ Lett.\  {\bf 97} (2006) 221803
  [arXiv:hep-ex/0608006];
  B.~Aubert {\it et al.}  [BABAR Collaboration],
  Phys.\ Rev.\ Lett.\  {\bf 98} (2007) 211802
  [arXiv:hep-ex/0703020];
  K.~Abe {\it et al.}  [BELLE Collaboration],
  arXiv:0704.1000 [hep-ex];
  M.~Staric {\it et al.}  [Belle Collaboration],
  Phys.\ Rev.\ Lett.\  {\bf 98} (2007) 211803
  [arXiv:hep-ex/0703036];
  B.~Aubert {\it et al.}  [BABAR Collaboration],
  Phys.\ Rev.\  D {\bf 76} (2007) 014018
  [arXiv:0705.0704 [hep-ex]].

\bibitem{cfactors}
  J.~M.~Soares and L.~Wolfenstein,
  Phys.\ Rev.\  D {\bf 47} (1993) 1021;
  N.~G.~Deshpande, B.~Dutta and S.~Oh,
  Phys.\ Rev.\ Lett.\  {\bf 77} (1996) 4499
  [arXiv:hep-ph/9608231];
  J.~P.~Silva and L.~Wolfenstein,
  Phys.\ Rev.\  D {\bf 55} (1997) 5331
  [arXiv:hep-ph/9610208];
  A.~G.~Cohen, D.~B.~Kaplan, F.~Lepeintre and A.~E.~Nelson,
  Phys.\ Rev.\ Lett.\  {\bf 78} (1997) 2300
  [arXiv:hep-ph/9610252];
  Y.~Grossman, Y.~Nir and M.~P.~Worah,
  Phys.\ Lett.\  B {\bf 407} (1997) 307
  [arXiv:hep-ph/9704287].

\bibitem{UTfitSM06}
  M.~Bona {\it et al.}  [UTfit Collaboration],
  JHEP {\bf 0610} (2006) 081
  [arXiv:hep-ph/0606167].

\bibitem{otherBs}
  Z.~Ligeti, M.~Papucci and G.~Perez,
  Phys.\ Rev.\ Lett.\  {\bf 97} (2006) 101801
  [arXiv:hep-ph/0604112];
  P.~Ball and R.~Fleischer,
  Eur.\ Phys.\ J.\  C {\bf 48} (2006) 413
  [arXiv:hep-ph/0604249].

\bibitem{massinsertion}
  M.~Ciuchini, E.~Franco, A.~Masiero and L.~Silvestrini,
  Phys.\ Rev.\  D {\bf 67} (2003) 075016
  [Erratum-ibid.\  D {\bf 68} (2003) 079901]
  [arXiv:hep-ph/0212397];
  J.~Foster, K.~i.~Okumura and L.~Roszkowski,
  JHEP {\bf 0603} (2006) 044
  [arXiv:hep-ph/0510422].

\bibitem{valentina}
  M.~Ciuchini \textit{et al.}, in preparation.

\bibitem{bpark}
  A.~Donini, V.~Gimenez, L.~Giusti and G.~Martinelli,
  Phys.\ Lett.\  B {\bf 470} (1999) 233
  [arXiv:hep-lat/9910017];
  R.~Babich, N.~Garron, C.~Hoelbling, J.~Howard, L.~Lellouch and C.~Rebbi,
  Phys.\ Rev.\  D {\bf 74} (2006) 073009
  [arXiv:hep-lat/0605016];
  Y.~Nakamura {\it et al.}  [CP-PACS Collaboration],
  PoS {\bf LAT2006} (2006) 089
  [arXiv:hep-lat/0610075].

\bibitem{hep-lat/0110091}
  D.~Becirevic, V.~Gimenez, G.~Martinelli, M.~Papinutto and J.~Reyes,
  JHEP {\bf 0204} (2002) 025
  [arXiv:hep-lat/0110091].

\end{thebibliography}
\end{document}